\documentclass[11pt]{article}
\usepackage[left=0.6in,right=0.6in]{geometry}
\usepackage{empheq}
\usepackage{amssymb}
\usepackage{dsfont}
\usepackage{slashed}
\usepackage[pdfstartview=FitH,pdfpagemode=None]{hyperref}
\usepackage{color}
\usepackage{multirow}
\usepackage{mathtools}
\usepackage{stackrel}
\usepackage{hyperref}

\newcommand{\be}{\begin{equation}}
\newcommand{\ee}{\end{equation}}
\newcommand{\ba}{\begin{eqnarray}}
\newcommand{\ea}{\end{eqnarray}}
\newcommand{\bea}{\begin{eqnarray}}
\newcommand{\eea}{\end{eqnarray}}

\makeatletter\@addtoreset{equation}{section}\makeatother

\begin{document}

\begin{titlepage}
\hfill LCTP-18-13

\begin{center}

{\Large \textbf{Toward Precision Holography in Type IIA with Wilson Loops }}\\[4em]

{\small  Jerem\'ias Aguilera-Damia${}^1$,  Alberto Faraggi$^{2}$, Leopoldo A.~Pando Zayas${}^{3,4}$, Vimal Rathee${}^{3}$ and Guillermo A. Silva${}^{1,4}$}\\[3em]
%
%

${}^{1}$\emph{Instituto de F\'isica de La Plata - CONICET \& Departamento de F\'isica\\
 UNLP C.C. 67, 1900 La Plata, Argentina}\\[1em]

${}^{2}$\emph{Departamento de Ciencias Fisicas,  Facultad de Ciencias Exactas\\
Universidad Andres Bello,  Sazie 2212, Piso 7, Santiago, Chile}\\[1em]

${}^{3}$\emph{Leinweber Center for Theoretical Physics,  Randall Laboratory of Physics\\ The University of
Michigan,  Ann Arbor, MI 48109, USA}\\[1em]

${}^4$\emph{The Abdus Salam International Centre for Theoretical Physics\\ Strada Costiera 11,  34014 Trieste, Italy}\\[6em]

\abstract{
We study the one-loop effective action of certain classical type IIA string configurations in  $AdS_4\times \mathbb{CP}^3$. These configurations are dual to  Wilson loops in the $\mathcal{N}= 6\:$ $U(N)_k \times U(N)_{-k}$ Chern-Simons theory coupled to matter  whose expectation values are known via supersymmetric localization. We compute the one-loop effective actions using two methods: perturbative heat kernel techniques and full $\zeta$-function regularization. We find that the result of the perturbative heat kernel method matches the field theory prediction at the appropriate order while the $\zeta$-function approach seems to lead to a disagreement. We explore various issues that might be responsible for this state of affairs.  }

\vspace{0.5cm}
{\tt jeremiasadlp@gmail.com, alberto.faraggi@unab.cl, lpandoz@umich.edu, \\vimalr@umich.edu, silva@fisica.unlp.edu.ar}
\end{center}

\end{titlepage}

\newpage
\tableofcontents

\section{Introduction}

The AdS/CFT correspondence explicitly realizes the notion that certain field theories admit an equivalent description in terms of string theories. The most prominent and  precise examples of such equivalences are: (i)  the duality between ${\cal N}=4$ supersymmetric Yang-Mills in four dimensions and type IIB string theory on $AdS_5\times S^5$ and (ii) ${\cal N}=6$ supersymmetric Chern-Simos with gauge group $U(N)_k\times U(N)_{-k}$ coupled to matter in three dimensions and type IIA string theory on $AdS_4\times \mathbb{CP}^3$.

 The advent of localization techniques has provided a plethora of exact results relevant for the field theory sides of this correspondence, that is, for ${\cal N}=4$ SYM \cite{Pestun:2007rz} and for ABJM \cite{Kapustin:2009kz}. In the context of the AdS/CFT correspondence, it is then natural to extrapolate the exact field theory results to the regime where they could be directly compared with the supergravity and semiclassical approximations.  This approach was attempted very early on in the insightful work of Drukker, Gross and Tseytlin \cite{Drukker:2000ep}; it did not, however, led to a match with the field theory prediction. This discrepancy motivated much work \cite{Sakaguchi:2007ea,Kruczenski:2008zk,Kristjansen:2012nz,Buchbinder:2014nia} that largely confirmed the original discrepancy. A recent revival of this line of effort took place in  \cite{Forini:2015bgo,Faraggi:2016ekd} which considered, on the gravity side,  the one-loop effective actions corresponding to the  ratio of the expectation values of the $\frac{1}{4}$ to  the $\frac{1}{2}$ BPS Wilson loops.  Various groups have made important subsequent contributions to this question \cite{Forini:2017whz}  and recently a precise match has been described, for the ${\cal N}=4$ SYM case, \cite{Cagnazzo:2017sny} after imposing a diffeomorphism preserving cutoff.  

 In  this manuscript we take one step in the direction of extending some of the techniques developed thus far to the context of the   AdS/CFT pair $AdS_4\times \mathbb{CP}^3$/ABJM.  We hope that by turning our attention to the $AdS_4$/ABJM pair we can gather complementary information to the one already available and ultimately learn about string perturbation theory in curved backgrounds with Ramond-Ramond fluxes. There are, indeed, a number of exact results obtained via localization of the  ABJM theory  starting with the free energy of the theory on $S^3$   \cite{Kapustin:2009kz} but most importantly to us there are various exact results for supersymmetric Wilson loops  for the $\frac{1}{2}$ BPS \cite{Klemm:2012ii} and, more recently, for the $\frac{1}{6}$ BPS configuration \cite{Bianchi:2018bke}. We consider one-loop effective actions of string configurations dual to those  supersymmetric Wilson loops in ABJM. Our focus is in understanding some aspects of the picture of precision holography, that is, the matching of sub-leading corrections on the string theory side with the prediction of field theory. As the first step in attacking the ${\cal N}=6$ case, in this first work we provide all the details to set the wheels of precision holography in $AdS_4\times \mathbb{CP}^3$/ABJM with Wilson loops in motion.

The rest of the paper is organized as follows.  We briefly review the field theory status of the expectation values of the relevant Wilson loops in section \ref{Sec:FieldTheory}.  In section \ref{Sec:ClassString} we discuss the classical string configurations   and in section \ref{Sec:QuadFlu} we present the quadratic fluctuations. The string theory semiclassical one-loop effective action is equivalent to the computations of quotients of determinants. In section \ref{Sec:Perturbative} we consider the perturbative computation of determinants to first non-trivial order in the latitude angle $\theta_0$.  Section 6 tackles the computation of the one-loop effective actions using $\zeta$-function regularization. We conclude with some comments and open problems in section \ref{Sec:Zeta}. We relegate a number of more technical aspects to  a series of appendices, including: conventions \ref{App:conventions}, geometric data  \ref{app: Geometric Data},   an explicit discussion of regularity conditions for the gauge fields  \ref{app:regularity}, and details of the fermionic reduction \ref{App:Spinors}.

\section{The $\frac{1}{6}$-BPS Latitude Wilson Loop}\label{Sec:FieldTheory}


The ABJM theory is a three-dimensional Chern-Simons-matter theory with $U(N)_k\times U(N)_{k}$ gauge group where the subindices indicate the Chern-Simons level \cite{Aharony:2008ug}.  The matter sector contains four complex scalar fields $C_I, (I=1,2,3,4)$ in the bifundamental representation $({\bf N}, \bar{\bf N})$ and the corresponding complex conjugate in the $(\bar{\bf N}, {\bf N})$ representation; the theory also contains fermionic superpartners  (see \cite{Aharony:2008ug}  for details). 

To build $\frac{1}{6}$ supersymmetric  Wilson loops, one starts considering only one of the gauge fields of the whole $U(N)\times U(N)$ gauge group, denoted by   $A_\mu$. To preserve supersymmetry we need to include a contribution from the matter sector. The main intuition comes from the construction of  supersymmetric Wilson loops in ${\cal N}=4$ SYM. However, in the absence of adjoint fields, an appropriate combination of bi-fundamentals, $C_I$, namely \cite{Drukker:2008zx,Rey:2008bh,Chen:2008bp} is required:
\be\label{Eq:WL16}
W_{\cal R}=\frac{1}{{\rm dim}[{\cal R}]}{\rm Tr}_{\cal R}\, {\cal P}\int \left(i A_\mu \dot{x}^\mu +\frac{2\pi}{k} |\dot{x}|\, M^I_J C_I \bar{C}^J\right) ds,
\ee
where ${\cal R}$ denotes the representation. It was shown in \cite{Drukker:2008zx,Rey:2008bh,Chen:2008bp} that the above operator preserves $\frac{1}{6}$ of the 24 supercharges when the loop is a straight line or a circle and the matrix takes the form $M^I_J={\rm diag}\, (1,1,-1,-1)$.

A remarkable result of \cite{Kapustin:2009kz} was to show that the computation of the vacuum expectation values of these Wilson loops reduces to a matrix model. Namely, the Wilson loop vev is obtained by inserting $\text{Tr}_{\cal R} \: e^{\mu_i}$ inside the following partition function:
\begin{equation}
Z(N,k) = \frac{1}{(N!)^2} \int\! \prod_{i=1}^N \frac{d\mu_i}{2\pi}\frac{d\nu_i}{2\pi}
\frac{\prod_{i<j}\left(2\sinh\frac{\mu_i-\mu_j}2\right)^2\left(2\sinh\frac{\nu_i-\nu_j}2\right)^2}{\prod_{i,j}\left(2\cosh\frac{\mu_i-\nu_j}2\right)^2}
\exp\!\left[\frac{ik}{4\pi}\sum_i (\mu_i^2-\nu_i^2)\right]\!.
\label{eq:ABJMZ}
\end{equation}
A particularly impressive exact result  was the computation of the supersymmetric free energy of ABJM on $S^3$ in terms of Airy functions \cite{Marino:2009jd,Drukker:2010nc}  which elucidated various aspects of the interpolation between week and strong coupling in the context of ABJM. The results that are more relevant for our current work pertain exact evaluations of Wilson loops.  The construction of the Wilson loop presented above  in equation (\ref{Eq:WL16}) does not capture the $\frac{1}{2}$ BPS string configurations. These involved the introduction of a superconnection  \cite{Drukker:2009hy}. The exact expectation values of the $\frac{1}{2}$- and certain $\frac{1}{6}$-BPS Wilson loops were presented in \cite{Klemm:2012ii} and take the general form 

\begin{equation}
\langle W^{\frac{1}{2}}_{\square}\rangle=\frac{1}{4}{\rm csc}\left(\frac{2\pi}{k}\right)\frac{\text{Ai} \big{[} \big{(} \frac{2}{\pi^2 k}\big{)}^{-1/3} \: \big{(} N - \frac{k}{24} - \frac{7}{3k} \big{)} \big{]}}{\text{Ai} \big{[} \big{(} \frac{2}{\pi^2 k}\big{)}^{-1/3} \: \big{(} N - \frac{k}{24} - \frac{1}{3k} \big{)} \big{]}},
\end{equation}
where the denominator is recognized as the partition function obtained in \cite{Fuji:2011km,Marino:2011eh}.  The above result and many others in this class are  exact to all orders in $1/N$,  up to exponentially small corrections in $N$. Recently, in \cite{Bianchi:2018bke}, a matrix model for the exact evaluation of the latitude BPS Wilson loops has been proposed. The expectation value for any genus of the  fermionic  (in the sense of the superconnection  \cite{Drukker:2009hy}) latitude Wilson loop is given in terms of Airy functions by (see equations (1.3) and (5.44) in \cite{Bianchi:2018bke}), 
\bea
\langle W_F^{\frac{1}{6}} (\nu) \rangle_{\nu}  = - \frac{\nu \: \Gamma (- \frac{\nu}{2}) \: \text{Ai} \big{[} \big{(} \frac{2}{\pi^2 k}\big{)}^{-1/3} \: \big{(} N - \frac{k}{24} - \frac{6 \nu + 1}{3k} \big{)} \big{]} }{ 2^{\nu + 2} \: \sqrt{\pi} \: \Gamma \big{(} \frac{3- \nu}{2} \big{)} \: \sin \big{(} \frac{2 \pi \nu}{k}\big{)} \: \text{Ai} \big{[} \big{(} \frac{2}{\pi^2 k}\big{)}^{-1/3} \: \big{(} N - \frac{k}{24} - \frac{1}{3k} \big{)} \big{]}},
\eea
where $\nu = \sin (2 \alpha) \: \cos \theta_0$, the angle $\alpha$ can be freely chosen and determines the coupling to matter, the geometric parameter we are interested in is $\theta_0$, and  $0 \leq \nu \leq 1$.  The beautiful result above is the culmination of an impressive series of papers \cite{Cardinali:2012ru,Bianchi:2014laa,Bonini:2016fnc,Bianchi:2017svd} (see also\cite{Bianchi:2017ozk,Bianchi:2018scb}).

The fermionic latitude Wilson loop maps to a type IIA string configuration in the $AdS_4 \times \mathbb{CP}^3$ background with endpoints moving in a circle inside $\mathbb{CP}^3$.  When expanded to the regime of validity of the holographic  computation, namely, taking the  leading genus-zero expansion in the above, it has been shown to coincide with the semi-classical string computation of the $\frac{1}{6}$-BPS Wilson loop expectation value \cite{Correa:2014aga}. 
\bea
\langle W_F^{\frac{1}{6}} (\nu) \rangle_{\nu} \vert_{g = 0} = - \iota \frac{2^{-\nu - 2} \: \kappa^{\nu} \: \Gamma \big{(} - \frac{\nu}{2}\big{)} }{\sqrt{\pi} \: \Gamma \big{(} \frac{3}{2} - \frac{\nu}{2} \big{)} }
\eea
We will consider the ratio of $\frac{1}{6}$-BPS Wilson loop expectation value with the $\frac{1}{2}$-BPS one, dual to a circular Wilson loop. Therefore, the field theory prediction to be matched to our computation of the one-loop effective action of the string configuration takes the form ($\nu = \cos \theta_0$)
\bea\begin{split}
 \Delta \Gamma^{\text{1-loop}}_{\text{effective}} (\theta_0) = \text{ln} \bigg{[} \frac{\langle W_F^{\frac{1}{6}} (\nu) \rangle_{\nu}}{\langle W_F^{\frac{1}{2}} (1) \rangle_{1}} \bigg{]} &=  \ln \bigg{(}  \frac{1}{\pi} \:\cot \bigg{(} \pi \frac{\cos \theta_0}{2} \bigg{)} \bigg{)}  - \ln \: \bigg{(} \sin^2 \frac{\theta_0}{2} \bigg{)} \\
 & + 2 \: \ln \: \Gamma \bigg{(} \cos^2 \frac{\theta_0}{2} \bigg{)}   - \ln \big{(} \Gamma (\cos \theta_0)\big{)}  -  \ln \big{(} \cos \theta_0\big{)} \\
 &= \frac{1}{2} \: \theta_0^2 + O (\theta_0^4).
\end{split}\eea
Anticipating the use of a perturbative result using heat kernel techniques, in the last line above we have expanded the field theory answer for small latitude angle $\theta_0$.

\section{String Configurations Dual to Supersymmetric Wilson Loops}\label{Sec:ClassString}
In this section we review the classical string configurations dual to the fermionic latitude family of BPS Wilson loops. We present these results for the convenience of the reader and to set up our notation but refer the interested reader to the original literature  \cite{Drukker:2008zx,Rey:2008bh,Chen:2008bp} for the $\frac{1}{2}$ BPS cofiguration and  \cite{Correa:2014aga} for the latitude $\frac{1}{6}$ BPS configuration.

\subsection{The $AdS_4\times\mathds{CP}^3$ background}
The Euclidean $AdS_4$ ($EAdS_4$) metric is written as an $\mathds{H}_2\times S^1$ foliation,
\begin{empheq}{align}\label{AdS5 metric}
	ds^2_{EAdS_4}&=\cosh^2u\left(\sinh^2\rho\,d\psi^2+d\rho^2\right)+\sinh^2u\,d\phi^2+du^2\,.
\end{empheq}
Similarly, the metric on $\mathds{CP}^3$ is taken to be
\begin{empheq}{align}
	ds^2_{\mathds{CP}^3}&=\frac{1}{4}\left[d\alpha^2+\cos^2\frac{\alpha}{2}\left(d\vartheta_1^2+\sin^2\vartheta_1\,d\varphi_1^2\right)+\sin^2\frac{\alpha}{2}\left(d\vartheta_2^2+\sin^2\vartheta_2\,d\varphi_2^2\right)\right.
	\nonumber\\
	&\phantom{=}\left.+\cos^2\frac{\alpha}{2}\sin^2\frac{\alpha}{2}\left(d\chi+\cos\vartheta_1\,d\varphi_1-\cos\vartheta_2\,d\varphi_2\right)^2\right]\,.
\end{empheq}
The full metric is
\begin{empheq}{alignat=7}
	ds^2&=L^2\left(ds^2_{EAdS_4}+4\,ds^2_{\mathds{CP}^3}\right)\,,
	&\qquad
	L^2&=\frac{R^3}{4k}\,.
\end{empheq}
Finally, the other background fields read
\begin{empheq}{alignat=7}
	e^{\Phi}&=\frac{2L}{k}\,,
	&\qquad
	F_{(4)}&=-\frac{3ikL^2}{2}\textrm{vol}\left(AdS_4\right)\,,
	&\qquad
	F_{(2)}&=\frac{k}{4}dA\,,
\end{empheq}
where
\begin{empheq}{alignat=7}
	\textrm{vol}\left(AdS_4\right)&=\cosh^2u\sinh u\sinh\rho\,d\psi\wedge d\rho\wedge du\wedge d\phi\,,
	\\
	A&=\cos\alpha\,d\chi+2\cos^2\frac{\alpha}{2}\cos\vartheta_1\,d\varphi_1+2\sin^2\frac{\alpha}{2}\cos\vartheta_2\,d\varphi_2\,.
\end{empheq}
The factor of $i$ in $F_{(4)}$ is due to the Euclidean continuation. The 2-form is proportional to the Kahler form in $\mathbb{CP}^3$.

\subsection{Classical String Solution}

The classical $1/6$-BPS string solution we are interested in has
\begin{empheq}{alignat=7}
\begin{split}\label{eq45}
	u&=0\,,
	\\
	\alpha&=0\,,
\end{split}
\qquad
\begin{split}
	\rho'&=-\sinh\rho\,,
	\\
	\vartheta_1'&=-\sin\vartheta_1\,,
\end{split}
\qquad
\begin{split}
	\psi&=\tau\,,
	\\
	\varphi_1&=\tau\,.
\end{split}
\end{empheq}
The induced metric is then
\begin{empheq}{alignat=7}
	ds^2&=L^2A\left(d\tau^2+d\sigma^2\right)\,,
	&\qquad
	A&=\sinh^2\rho+\sin^2\vartheta_1^2=\rho'^2+\vartheta_1'^2\,.
\end{empheq}
The solution to \eqref{eq45} involves the latitude parameter $\theta_0$. We write,
\begin{empheq}{alignat=7}
	\sinh\rho&=\frac{1}{\sinh\sigma}\,,
	&\qquad
	\sin\vartheta_1&=\frac{1}{\cosh\left(\sigma+\sigma_0\right)}\,,
	&\qquad
	\cos \theta_0 &= \tanh \sigma_0.
\end{empheq}
The induced geometry is disk shaped and asymptotes $AdS_2$ at the boundary. The $1/2$-BPS limit corresponds to $\sigma_0\rightarrow\infty$, and the induced geometry then becomes an exact $AdS_2$.


\subsection{Symmetries of the classical solution}\label{subsec: symmetries}


We start by recalling  that the background geometry is constructed out from coset spaces $AdS_4=SO(2,3)/SO(1,3)$ and $\mathbb{CP}^3=SU(4)/SU(3)\times U(1)$.

Before gauge-fixing, the string embedding is characterized by $10$ worldsheet scalars $x^m(\tau,\sigma)$ and a 10-dimensional Majorana spinor $\theta$ whose dynamics is determined by the type IIA Green-Schwarz action (more details below). The symmetries of the theory are:
\begin{itemize}
	\item Local:
	\begin{itemize}
		\item Diffeomorphisms:
		\begin{empheq}{alignat=7}\label{eq: local symmetry}
			\delta_{\xi}x^m&=\xi^a\partial_ax^m\,,
			&\qquad
			\delta_{\xi}\theta&=\xi^a\partial_a\theta\,,
		\end{empheq}
		where $\xi^a$ is an arbitrary worldsheet vector field.
		\item $\kappa$-symmetry:
		\begin{empheq}{alignat=7}
			\delta_{\kappa}x^m&=\frac{i}{2}\overline{\theta}\Gamma^m\delta_{\kappa}\theta\,,
			&\qquad
			\delta_{\kappa}\theta&=\left(1+\Gamma_F\right)\kappa\,,
			&\qquad
			\Gamma_F&=\frac{\epsilon^{ab}}{2\sqrt{-g}}\Gamma_{ab}\Gamma_{11}\,,
		\end{empheq}
		where $\kappa$ is an arbitrary 10-dimensional Majorana spinor and worldsheet scalar.
	\end{itemize}
	\item Global:
	\begin{itemize}
		\item Target space isometries:
		\begin{empheq}{alignat=7}\label{eq: global symmetry}
			\delta_{\lambda}x^m&=K^m\,,
			&\qquad
			\delta_{\lambda}\theta&=K^a\partial_a\theta-\frac{1}{4}\left(\nabla_mK_n-\nabla_nK_m\right)\Gamma^{mn}\theta\,,
		\end{empheq}
		where $K^m$ is any target space Killing vector and $K_a=\partial_ax^mK_m$.
		\item Target space supersymmetries:
		\begin{empheq}{alignat=7}
			\delta_{\epsilon}x^m&=-\frac{i}{2}\overline{\theta}\Gamma^m\delta_{\epsilon}\theta\,,
			&\qquad
			\delta_{\epsilon}\theta&=\epsilon\,,
			&\qquad
			D_m\epsilon&=0\,,
	\end{empheq}
	where $\epsilon$ is any target space Killing spinor.
	\end{itemize}
\end{itemize}

Given a classical solution (with fermions set to zero, $\theta=0$), the preserved bosonic symmetries correspond to the set of transformations satisfying
\begin{empheq}{alignat=7}
	\delta x^m&=0
	&\qquad\Rightarrow\qquad
	K^m+\epsilon^a\partial_ax^m&=0\,.
\end{empheq}
In other words, the target space isometries inherited by the solution are those that leave the embedding invariant up to worldsheet diffeomorphisms. Contracting this condition with $g_{mn}\partial_ax^n$ we can solve
\begin{empheq}{alignat=7}
	\epsilon^a&=-K^a\,,
\end{empheq}
where $K_a=\partial_ax^mK_m$. This in turn implies that, in order to generate a symmetry, the Killing vector must satisfy
\begin{empheq}{alignat=7}
	K^m&=g^{ab}\partial_ax^m\partial_bx^nK_n\,.
\end{empheq}

The logic for the fermionic symmetries is the same. The ones preserved by the background are those satisfying
\begin{empheq}{alignat=7}
	\delta\theta&=0
	&\qquad\Rightarrow\qquad
	\epsilon+\left(1+\Gamma_F\right)\kappa&=0\,.
\end{empheq}
These are target space supersymmetries which can be compensated by a local $\kappa$-symmetry transformation. Multiplying by $\left(1-\Gamma_F\right)$, we find that
\begin{empheq}{alignat=7}
	\left(1-\Gamma_F\right)\epsilon&=0\,.
\end{empheq}
This is the usual condition for preserved supersymmetries. This condition is in fact sufficient since then we can solve
\begin{empheq}{alignat=7}
	\kappa&=-\frac{1}{2}\epsilon\,.
\end{empheq}

For the case at hand, we find that the $AdS_4\times\mathds{CP}^3$ Killing vectors
\begin{empheq}{alignat=7}\label{eq: Killing vectors}
	K_1&=\partial_{\psi}+\partial_{\varphi_1}\,,
	\\\nonumber
	K_2&=\partial_{\phi}\,,
	\\\nonumber
	K_3&=-\cos\varphi_2\,\partial_{\vartheta_2}+\cot\vartheta_2\sin\varphi_2\,\partial_{\varphi_2}+\frac{\sin\varphi_2}{\sin\vartheta_2}\partial_{\chi}\,,
	\\\nonumber
	K_4&=\sin\varphi_2\,\partial_{\vartheta_2}+\cot\vartheta_2\cos\varphi_2\,\partial_{\varphi_2}+	\frac{\cos\varphi_2}{\sin\vartheta_2}\partial_{\chi}\,,
	\\\nonumber
	K_5&=\partial_{\varphi_2}
	\\\nonumber
	K_6&=\partial_{\chi}\,,
\end{empheq}
generate a symmetry of the solution. The first Killing vector must be accompanied by a translation in the worldsheet coordinate $\tau$ such that $\epsilon^{\tau}_{cl}=-\lambda_{cl}$ and $\epsilon^{\sigma}_{cl}=0$; it corresponds to an isometry of the induced geometry. The rest have zero norm on the worldsheet so $\epsilon^a_{cl}=0$. Altogether we have a $\underbrace{U(1)}_{K_1}\times\underbrace{U(1)}_{K_2}\times\underbrace{SU(2)}_{K_3,K_4,K_5}\times\underbrace{U(1)}_{K_6}$ symmetry.

The geometric interpretation of the symmetries is most easily seen in the embedding coordinates of $EAdS_4\subset\mathds{R}^5$ and the Hopf fibration $S^1\hookrightarrow S^7\rightarrow\mathds{CP}^3$:
\begin{empheq}{alignat=7}
\begin{split}
	1&=X_0^2-X_1^2-X_2^2-X_3^2-X_4^2\,,
	\\
	ds^2&=-dX_0^2+dX_1^2+dX_2^2+dX_3^2+dX_4^2\,,
\end{split}
\qquad
\begin{split}
	X_0&=\cosh u\cosh\rho\,,
	\\
	X_1&=\cosh u\sinh\rho\cos\psi\,,
	\\
	X_2&=\cosh u\sinh\rho\sin\psi\,,
	\\
	X_3&=\sinh u\cos\phi\,,
	\\
	X_4&=\sinh u\sin\phi\,,
\end{split}
\end{empheq}
\begin{empheq}{alignat=7}
	z_1&=\cos\frac{\alpha}{2}\cos\frac{\vartheta_1}{2}e^{\frac{i}{2}\left(\varphi_1+\frac{\chi}{2}\right)}\,,
	&\qquad
	z_3&=\sin\frac{\alpha}{2}\cos\frac{\vartheta_2}{2}e^{\frac{i}{2}\left(\varphi_2-\frac{\chi}{2}\right)}\,,
	\\
	z_2&=\cos\frac{\alpha}{2}\sin\frac{\vartheta_1}{2}e^{\frac{i}{2}\left(-\varphi_1+\frac{\chi}{2}\right)}\,,
	&\qquad
	z_4&=\sin\frac{\alpha}{2}\sin\frac{\vartheta_2}{2}e^{\frac{i}{2}\left(-\varphi_2-\frac{\chi}{2}\right)}\,.
\end{empheq}
The worldsheet has $z_3=z_4=0$.

In the next section we will consider perturbations of the string embedding around the classical solution and look at the transformation properties of the fluctuations under the preserved symmetries. It will prove convenient to take linear combinations of $K_3$, $K_4$ and $K_5$ that have a simple action on the fluctuations. We find that such combinations are
\begin{empheq}{alignat=7}\label{eq: preserved generators}
	K_3'&=\cos(\vartheta_2^{cl})\left(\sin(\varphi_2^{cl})K_3+\cos(\varphi_2^{cl})K_4\right)+\sin(\vartheta_2^{cl})K_5\,,
	\\
	K_4'&=\cos(\varphi_2^{cl})K_3-\sin(\varphi_2^{cl})K_4\,,
	\\
	K_5'&=\sin(\vartheta_2^{cl})\left(\sin(\varphi_2^{cl})K_3+\cos(\varphi_2^{cl})K_4\right)-\cos(\vartheta_2^{cl})K_5\,,
\end{empheq}
\begin{empheq}{alignat*=7}
	K_3'&=\cos(\vartheta_2^{cl})\sin(\varphi_2-\varphi_2^{cl})\partial_{\vartheta_2}+\left(\cot\vartheta_2\cos(\vartheta_2^{cl})\cos(\varphi_2-\varphi_2^{cl})+\sin\vartheta_2^{cl}\right)\partial_{\varphi_2}+\frac{\cos(\vartheta_2^{cl})\cos(\varphi_2-\varphi_2^{cl})}{\sin\vartheta_2}\partial_{\chi}\,,
\end{empheq}
where $\vartheta_2^{cl}$ and $\varphi_2^{cl}$ are the (constant) values that the coordinates $\vartheta_2$ and $\varphi_2$ take on the classical solution. We shall drop the primes henceforth.

\section{Quadratic Fluctuations}\label{Sec:QuadFlu}

Having reviewed the classical solution dual to the $\frac{1}{6}$-BPS latitude Wilson loop and its symmetries, in this section we derive the corresponding spectrum of quadratic fluctuations. There has already been some previous work for the case of the $\frac{1}{2}$-BPS configuration in \cite{Sakaguchi:2010dg} and \cite{Kim:2012nd}  whose spectrum is a limit of our result. We will start by giving a general expression for the quadratic fluctuations of the type IIA string in $AdS_4\times\mathds{CP}^3$ and then specialize to the case of the $\frac{1}{6}$ BPS string dual to the latitude Wilson loop. In what follows, target-space indices are denoted by $m,n,\ldots$, world-sheet indices are $a,b,\ldots$, while the directions orthogonal to the string are represented by $i,j,\ldots$. All corresponding tangent space indices are underlined.
\subsection{Type IIA strings on $AdS_4\times\mathds{CP}^3$}
In the bosonic sector, the string dynamics is dictated by the Nambu-Goto (NG) action
\begin{empheq}{align}
	S_\textrm{NG}&=\frac{1}{2\pi \alpha'}\int d^2\sigma\,\sqrt{-g}\,,
\end{empheq}
where $g_{ab}$ is the induced metric on the world sheet and $g=\det g_{ab}$. Our first goal in this section is to consider perturbations $x^m\rightarrow x^m+\varepsilon y^m$, $\varepsilon\ll1$, around any given classical embedding and find the quadratic action that governs them. To this purpose, let us choose convenient vielbeins for the $AdS_4\times\mathds{CP}^3$ metric that are properly adapted to the study of fluctuations. Using the local $SO(9,1)$ symmetry, we can always pick a frame $E^{\underline{m}}=(E^{\underline{a}},E^{\underline{i}})$ such that the pullback of $E^{\underline{a}}$ onto the world-sheet forms a vielbein for the induced metric, while the pullback of $E^{\underline{i}}$ vanishes. Of course, these are nothing but the 1-forms dual to the tanget and normal vectors fields, respectively. The Lorentz symmetry is consequently broken to $SO(1,1)\times SO(8)$. Having made this choice we may define the fields
\begin{empheq}{align}
	\chi^{\underline{m}}&\equiv E^{\underline{m}}_{\phantom{\underline{m}}m}y^m\,,
\end{empheq}
and gauge fix the diffeomorphism invariance by freezing the tangent fluctuations, namely, by requiring
\begin{empheq}{align}
	\chi^{\underline{a}}&=0\,.
\end{empheq}
The physical degrees of freedom are then parameterized by the normal directions $\chi^{\underline{i}}$. In this gauge the variation of the induced metric is
\begin{empheq}{align}
	\varepsilon^{-1}\delta g_{ab}&=-2H_{\underline{i}ab}\chi^{\underline{i}}+
	\nabla_a\chi^{\underline{i}}\nabla_b\chi^{\underline{j}}\delta_{\underline{ij}}
	+\left(H_{\underline{i}a}^{\phantom{\underline{i}a}c}H_{\underline{j}bc}
	-R_{m\underline{i}n\underline{j}}\partial_ax^m\partial_bx^n\right)\chi^{\underline{i}}\chi^{\underline{j}}\,,
\end{empheq}
where $H^{\underline{i}}_{\phantom{\underline{i}}ab}$ is the extrinsic curvature of the embedding and
\begin{empheq}{alignat=5}
	 \nabla_a\chi^{\underline{i}}&=\partial_a\chi^{\underline{i}}+\mathcal{A}^{\underline{ij}}_{\phantom{\underline{ij}}a}\chi_j
\end{empheq}
is the world-sheet covariant derivative, which includes the $SO(8)$ normal bundle connection $\mathcal{A}^{\underline{ij}}_{\phantom{\underline{ij}}a}$. These objects, as well as the world-sheet spin connection $w^{\underline{ab}}$, are related to the pullback of the target-space spin connection $\Omega^{\underline{mn}}$ by
\begin{empheq}{alignat=5}
	w^{\underline{ab}}&=P[\Omega^{\underline{ab}}]\,,
	&\qquad
	 H^{\underline{i}}_{\phantom{\underline{i}}ab}&=P[\Omega^{\underline{i}}_{\phantom{\underline{i}}\underline{a}}]_ae^{\underline{a}}_{\phantom{\underline{a}}b}\,,
	&\qquad
	\mathcal{A}^{\underline{ij}}&=P[\Omega^{\underline{ij}}]\,,
\end{empheq}
where $e^{\underline{a}}_{\phantom{\underline{a}}a}=P\left[E^{\underline{a}}\right]_a$ is the induced geometry vielbein. Using the well-known expansion of the square root of a determinant, a short calculation shows that, to quadratic order, the NG action becomes
\begin{empheq}{align}
	S^{(2)}_\textrm{NG}&=\frac{1}{4\pi\alpha'}\int d\tau d\sigma\,\sqrt{-g}\left(g^{ab}\nabla_a\chi^{\underline{i}}\nabla_b\chi^{\underline{j}}\delta_{\underline{ij}}-\left(g^{ab}H_{\underline{i}a}^{\phantom{\underline{i}a}c}H_{\underline{j}bc}+\delta^{\underline{ab}}R_{\underline{aibj}}\right)\chi^{\underline{i}}\chi^{\underline{j}}\right)\,,
\end{empheq}
where we have used the equations of motion $g^{ab}H^{\underline{i}}_{\phantom{\underline{i}}ab}=0$ and written $g^{ab}R_{m\underline{i}n\underline{j}}\partial_ax^m\partial_bx^n=\delta^{\underline{ab}}R_{\underline{aibj}}$. The continuation of this expression to Euclidean signature is straightforward.

Let us now discuss the fermionic degrees of freedom. In Lorentzian signature, the type IIA string involves a single 10-dimensional Majorana spinor $\theta$. At quadratic order, the Green-Schwarz (GS) action that controls its dynamics on $AdS_4\times\mathds{CP}^3$ is given by
\begin{empheq}{alignat=7}
	S_{\textrm{GS}}&=\frac{i}{4\pi\alpha'}\int d^2\sigma\sqrt{-g}\,\overline{\theta}\left(g^{ab}-\frac{\epsilon^{ab}}{\sqrt{-g}}\Gamma_{11}\right)\Gamma_aD_b\theta\,,
\end{empheq}
where the symbol $\epsilon^{ab}$ is a density with $\epsilon^{\tau\sigma}=1$, $\Gamma_a=\Gamma_m\partial_ax^m$ is the pullback of the 10-dimensional Dirac matrices and $\Gamma_{11}\equiv\Gamma_{\underline{0123456789}}$. Also, $D_a=\partial_ax^mD_m$ is the pullback of the spacetime covariant derivative appearing in the supersymmetry variation of the gravitino, which includes the contribution from the RR fluxes. Explicitly,
\begin{empheq}{alignat=7}
	D_a&=\partial_ax^m\nabla_m+\frac{1}{8}e^{\Phi}\left[\slashed{F}_{(2)}\Gamma_{11}+\slashed{F}_{(4)}\right]\Gamma_a\,.
\end{empheq}

The above action can be simplified considerably. Indeed, given our choice of vielbein we have
\begin{empheq}{align}
	 D_a&=\nabla_a-\frac{1}{2}H^{\underline{i}\phantom{a}\underline{a}}_{\phantom{\underline{i}}a}\Gamma_{\underline{ai}}+\frac{1}{8}e^{\Phi}\left[\slashed{F}_{(2)}\Gamma_{11}+\slashed{F}_{(4)}\right]\Gamma_a\,,
\end{empheq}
where the world-sheet covariant derivative $\nabla_a$ includes the normal bundle connection $\mathcal{A}^{\underline{ij}}_{\phantom{\underline{ij}}a}$, that is,
\begin{empheq}{align}
	 \nabla_a&=\partial_a+\frac{1}{4}w^{\underline{ab}}_{\phantom{\underline{ab}}a}\Gamma_{\underline{ab}}+\frac{1}{4}\mathcal{A}^{\underline{ij}}_{\phantom{\underline{ij}}a}\Gamma_{\underline{ij}}\,.
\end{empheq}
Using the relation $\epsilon^{ab}\Gamma_a=\sqrt{-g}\,\Gamma_{\underline{01}}\Gamma^b$, it is easy to see that the terms proportional to the extrinsic curvature drop out from the action because of the equations of motion $H^{\underline{i}}_{\phantom{\underline{i}}ab}\Gamma^a\Gamma^b=H^{\underline{i}}_{\phantom{\underline{i}}ab}g^{ab}=0$. Then,
\begin{empheq}{alignat=5}
	S_{\textrm{GS}}&=\frac{i}{4\pi\alpha'}\int d\tau d\sigma\,\sqrt{-g}\,\overline{\theta}\left(1-\Gamma_{\underline{01}}\Gamma_{11}\right)\Gamma^a\left(\nabla_a+\frac{1}{8}e^{\Phi}\left[\slashed{F}_{(2)}\Gamma_{11}+\slashed{F}_{(4)}\right]\Gamma_a\right)\theta\,.
\end{empheq}
Now, in addition to diffeomorphism invariance and local Lorentz rotations, the full string action enjoys the local $\kappa$-symmetry
\begin{empheq}{alignat=5}
	\delta_{\kappa}\theta&=\frac{1}{2}\left(1+\Gamma_{\underline{01}}\Gamma_{11}\right)\kappa\,,
	&\qquad
	\delta_{\kappa}x^m&=\frac{i}{2}\overline{\theta}\Gamma^m\delta_{\kappa}\theta\,.
\end{empheq}
It is then possible to gauge fix to
\begin{empheq}{alignat=7}
	\frac{1}{2}\left(1-\Gamma_{\underline{01}}\Gamma_{11}\right)\theta&=\theta
	&\qquad\Leftrightarrow\qquad
	\frac{1}{2}\overline{\theta}\left(1-\Gamma_{\underline{01}}\Gamma_{11}\right)&=\overline{\theta}\,,
\end{empheq}
resulting in
\begin{empheq}{alignat=5}
	S_{\textrm{GS}}&=\frac{i}{2\pi\alpha'}\int d\tau d\sigma\,\sqrt{-g}\,\overline{\theta}\,\Gamma^a\left(\nabla_a+\frac{1}{8}e^{\Phi}\left[-\slashed{F}_{(2)}\Gamma_{\underline{01}}+\slashed{F}_{(4)}\right]\Gamma_a\right)\theta\,.
\end{empheq}
Finally, we will need the Euclidean continuation of the action:
\begin{empheq}{alignat=7}
	S_{\textrm{GS}}&=\frac{1}{2\pi\alpha'}\int d\tau d\sigma\,\sqrt{g}\,\overline{\theta}\,\Gamma^a\left(\nabla_a+\frac{1}{8}e^{\Phi}\left(i\slashed{F}_{(2)}\Gamma_{\underline{01}}+\slashed{F}_{(4)}\right)\Gamma_a\right)\theta\,.
\end{empheq}
The $\kappa$-symmetry fixing becomes $i\Gamma_{\underline{01}}\Gamma_{11}\theta=\theta$ where now $\Gamma_{11}\equiv-i\Gamma_{\underline{0123456789}}$. We will take this expression as our starting point; all quantities involved are intrinsically Euclidean, including the fluxes and Dirac matrices.

\subsection{Bosonic Fluctuations}

Putting everything together we find that the action that governs the bosonic fluctuations is
\begin{empheq}{alignat=7}
	S_{(2,3)}&=\frac{L^2}{\pi\alpha'}\int d\tau d\sigma\sqrt{g}\left(g^{ab}\left(\partial_a\chi^{\underline{23}}\right)^*\partial_b\chi^{\underline{23}}+\frac{2\sinh^2\rho}{\sqrt{g}}\left|\chi^{\underline{23}}\right|^2\right)\,,
	&\qquad
	\chi^{\underline{23}}&=\frac{1}{\sqrt{2}}\left(\chi^{\underline{2}}+i\chi^{\underline{3}}\right)\,,
	\\
	S_{(4,5)}&=\frac{L^2}{\pi\alpha'}\int d\tau d\sigma\sqrt{g}\left(g^{ab}\left(D^{\mathcal{A}}_a\chi^{\underline{45}}\right)^*D^{\mathcal{A}}_b\chi^{\underline{45}}-\frac{2m^2}{\sqrt{g}}\left|\chi^{\underline{45}}\right|^2\right)\,,
	&\qquad
	\chi^{\underline{45}}&=\frac{1}{\sqrt{2}}\left(\chi^{\underline{4}}+i\chi^{\underline{5}}\right)\,,
	\\
	S_{(6,7)}&=\frac{L^2}{\pi\alpha'}\int d\tau d\sigma\sqrt{g}\left(g^{ab}\left(D^{\mathcal{B}}_a\chi^{\underline{67}}\right)^*D^{\mathcal{B}}_b\chi^{\underline{67}}-\frac{\sin^2\vartheta_1}{2\sqrt{g}}\left|\chi^{\underline{67}}\right|^2\right)\,,
	&\qquad
	\chi^{\underline{67}}&=\frac{1}{\sqrt{2}}\left(\chi^{\underline{6}}+i\chi^{\underline{7}}\right)\,,
	\\
	S_{(8,9)}&=\frac{L^2}{\pi\alpha'}\int d\tau d\sigma\sqrt{g}\left(g^{ab}\left(D^{\mathcal{B}}_a\chi^{\underline{89}}\right)^*D^{\mathcal{B}}_b\chi^{\underline{89}}-\frac{\sin^2\vartheta_1}{2\sqrt{g}}\left|\chi^{\underline{89}}\right|^2\right)\,,
	&\qquad
	\chi^{\underline{89}}&=\frac{1}{\sqrt{2}}\left(\chi^{\underline{8}}+i\chi^{\underline{9}}\right)\,,
\end{empheq}
where 
\begin{empheq}{alignat=7}
	m&=\frac{\sinh\rho\sin\vartheta_1}{\cosh\rho-\cos\vartheta_1},
\end{empheq}
and the $U(1)$ covariant derivatives read
\begin{empheq}{alignat=7}
	D^{\mathcal{A}}&=d+i\mathcal{A}\,,
	&\qquad
	D^{\mathcal{B}}&=d+i\mathcal{B}\,,
\end{empheq}
with
\begin{empheq}{alignat=7}\label{Eq:Connections}
	\mathcal{A}&\equiv\mathcal{A}^{\underline{45}}=\left(1 - \frac{\cosh\rho\cos\vartheta_1+1}{\cosh\rho+\cos\vartheta_1}\right)d\tau, 
	\\
	\mathcal{B}&\equiv\mathcal{A}^{\underline{67}}=\mathcal{A}^{\underline{89}}=\frac{1}{2}\left(\cos\vartheta_1 - 1 \right)d\tau. \nonumber 
\end{empheq}
We have factored out the radius $L$ from the metric and the fluctuations. Notice that the $U(1)\times U(1)\times SU(2)\times U(1)$ symmetry structure is evident, with $\chi^{\underline{67}}$ and $\chi^{\underline{89}}$ forming a doublet.

\subsection{Fermionic Fluctuations}
For the case at hand, the fermionic action reads
\begin{empheq}{alignat=7}
	S_{\textrm{GS}}&=\frac{L^2}{\pi\alpha'}\int d\tau d\sigma\,\sqrt{g}\,\overline{\theta}\left(\Gamma^a\nabla_a+M\right)\theta\,,
\end{empheq}
where
\begin{empheq}{alignat=7}
	\nabla_{\tau}&=\partial_{\tau}+\frac{1}{2}\Gamma^{\underline{01}}\,w+\frac{1}{2}\Gamma^{\underline{45}}\,\mathcal{A}+\frac{1}{2}\left(\Gamma^{\underline{67}}+\Gamma^{\underline{89}}\right)\mathcal{B}\,,
	\\
	\nabla_{\sigma}&=\partial_{\sigma}\,,
	\\
	M&=\frac{i\Gamma^{\underline{01}}}{4A}\left(\left(3\Gamma^{\underline{23}}-\Gamma^{\underline{45}}\right)\left(\sinh^2\rho-\sin^2\vartheta_1\,\Gamma^{\underline{0145}}\right)+\left(\Gamma^{\underline{67}}+\Gamma^{\underline{89}}\right)A\right)\, ,
\end{empheq}
where ${\cal A}$ and ${\cal B}$ are the connections defined above in equation (\ref{Eq:Connections})  while $A$ is the conformal factor of the induced worldsheet metric and $w$ is defined in appendix \ref{app: Geometric Data}.

As for the bosons we have extracted the radius $L$ from the metric and rescaled the fermionic fields by $L^{1/2}$. The symmetry of the action under the $U(1)\times U(1)\times SU(2)\times U(1)$ bosonic subgroup follows from the fact that all the objects involved commute with the preserved generators \eqref{eq: preserved generators}.

\subsection{One-loop Effective Action}

The induced world-sheet geometry is that of the 2d Euclidean manifold $\mathcal{M}$ with the metric 
\bea\begin{split}
ds^2_{\mathcal{M}} &=  M (\rho) \: \big{(} d \rho^2 + \sinh^2 \rho \: d \tau^2  \big{)}, \\
M (\rho) &= 1 + \frac{\sin^2 \theta (\rho)}{\sinh^2 \rho}, \:\:\:\:\: \sin \theta (\rho) = \frac{\sinh \rho \: \sin \theta_0}{\cosh \rho + \cos \theta_0} 
\end{split}\eea
where $0 \leq \theta_0 \leq \frac{\pi}{2}$ is the latitude angle.  $\theta_0 = 0$ corresponds to the $\frac{1}{2}$- BPS solution.

The difference in 1-loop effective actions of $\frac{1}{6}$-BPS string withrespect to the $\frac{1}{2}$-BPS is
\bea\label{eq6}
e^{-\Delta \: \Gamma^{\text{1-loop}}_{\text{effective}} (\theta_0)} = \Bigg{[}\frac{  \Big{(} \frac{\text{det}\: \mathcal{O}_{4 +} (\theta_0)}{\text{det}\: \mathcal{O}_{4+} (0)}\Big{)} \: \Big{(} \frac{\text{det}\: \mathcal{O}_{4 -} (\theta_0)}{\text{det}\: \mathcal{O}_{4-} (0)}\Big{)} \: \Big{(} \frac{\text{det}\: \mathcal{O}_{5 +} (\theta_0)}{\text{det}\: \mathcal{O}_{5+} (0)}\Big{)}^3 \: \Big{(} \frac{\text{det}\: \mathcal{O}_{5 -} (\theta_0)}{\text{det}\: \mathcal{O}_{5-} (0)}\Big{)}^3 }{  \Big{(} \frac{\text{det}\: \mathcal{O}_{1} (\theta_0)}{\text{det}\: \mathcal{O}_{1} (0)}\Big{)}^2 \: \Big{(} \frac{\text{det}\: \mathcal{O}_{2 +} (\theta_0)}{\text{det}\: \mathcal{O}_{2+} (0)}\Big{)} \: \Big{(} \frac{\text{det}\: \mathcal{O}_{2 -} (\theta_0)}{\text{det}\: \mathcal{O}_{2-} (0)}\Big{)}\: \Big{(} \frac{\text{det}\: \mathcal{O}_{3 +} (\theta_0)}{\text{det}\: \mathcal{O}_{3+} (0)}\Big{)}^2 \: \Big{(} \frac{\text{det}\: \mathcal{O}_{3 -} (\theta_0)}{\text{det}\: \mathcal{O}_{3-} (0)}\Big{)}^2 } \Bigg{]}^{\frac{1}{2}}
\eea
where the bosonic spectrum of operators is
\bea\begin{split}
\mathcal{O}_1 (\theta_0) &= M^{-1} \big{(} -g^{\mu \nu} \: \nabla_{\mu} \nabla_{\nu} + 2 \big{)}, \\
\mathcal{O}_{2 \pm} (\theta_0) &= M^{-1} \big{(} -g^{\mu \nu} \: D^a_{\mu} \:D^a_{\nu} + V_2 \big{)}, \:\:\:\:\:\:\:\:\: D^a_{\mu} = \nabla_{\mu} \pm \iota \: \mathcal{A}_{\mu}, \\
\mathcal{O}_{3 \pm} (\theta_0) &= M^{-1} \big{(} -g^{\mu \nu} \: D^b_{\mu} \:D^b_{\nu} + V_3 \big{)}, \:\:\:\:\:\:\:\:\: D^b_{\mu} = \nabla_{\mu} \pm \iota \: \mathcal{B}_{\mu},\\
\end{split}\eea
and the fermionic operator is 
\bea\begin{split}
\mathcal{O}_{\alpha, \beta, \gamma} (\theta_0) &= M^{-\frac{1}{2}} \: \bigg{(} - \iota \: \bigg{(} \slashed{D} + \frac{1}{4} \slashed{\partial} \: \ln M \bigg{)} - \iota \: \Gamma_{\underline{01}} \: \big{(} m + V\big{)} + \alpha W \bigg{)}, 
\end{split}\eea
with $\mathcal{O}_{4, \alpha} $ in \eqref{eq6}, corresponding to the $-\alpha = \beta = \gamma$ case and $\mathcal{O}_{5, \alpha} $ corresponding to the $\alpha = \beta = \gamma$ and $\beta = - \gamma$ case. 

Here, we have $\mathcal{A}_{\rho} = \mathcal{B}_{\rho} = 0$, $\mathcal{A}_{\tau} = \mathcal{A} (\rho)$, $\mathcal{B}_{\tau} = \mathcal{B} (\rho)$  with $g_{\mu \nu}$ and $\nabla_{\mu}$ evaluated for the $AdS_2$ metric,
\bea
D_{\mu} = \nabla_{\mu} + \iota \frac{ \alpha }{2} \: \mathcal{A}_{\mu} + \iota \frac{ \beta + \gamma }{2} \: \mathcal{B}_{\mu},
\eea
and 
\bea
M (\rho) = 1 + \frac{\sin^2 \theta (\rho)}{\sinh^2 \rho}, \:\:\: \mathcal{A} (\rho) &=&  1 - \frac{1 + \cosh \rho \: \cos \theta (\rho)}{\cosh \rho + \cos \theta (\rho)}, \:\:\: \mathcal{B} (\rho) = \frac{1}{2} \big{(}  \cos \theta (\rho) - 1\big{)}  , \\
V_2 (\rho) &=& - \frac{\partial_{\rho}\mathcal{A} (\rho)}{\sinh \rho},  \:\:\: V_3 (\rho) =  - \frac{\partial_{\rho}\mathcal{B} (\rho)}{\sinh \rho}, \\
V(\rho) &=&  \frac{ (1 - 3 \: \beta \: \gamma)}{4} \: \frac{1}{\sqrt{M(\rho)}} - \frac{\alpha (\beta + \gamma) }{4} \: \sqrt{M (\rho)} - m, \\
W (\rho) &=& \frac{1 - 3 \:\beta \: \gamma }{4} \: \frac{\sin^2 \theta (\rho)}{\sqrt{M(\rho)} \: \sinh^2 \rho}. 
\eea
Here m corresponds to the value  of potential, V, at $\rho = \infty$.
\bea
m = \frac{ (1 - 3 \: \beta \: \gamma)}{4} - \frac{\alpha (\beta + \gamma) }{4} 
\label{mass}
\eea
The $\theta (\rho)$ dependence is given by,
\bea
\sin \theta (\rho) = \frac{\sinh \rho \: \sin \theta_0}{\cosh \rho + \cos \theta_0}, \:\:\:\:\: \cos \theta (\rho) = \frac{1 + \cosh \rho \: \cos \theta_0}{\cosh \rho + \cos \theta_0}.
\eea


\section{One-loop Effective Action: Perturbative Heat Kernel}\label{Sec:Perturbative}

We now proceed to evaluate fluctuations determinant using the heat kernel techniques. To evaluate the determinants we will exploit the fact that heat kernel techniques for $AdS_2$ are well-developed \cite{Camporesi:1994ga,Camporesi:1995fb,Buchbinder:2014nia}.  More precisely, we will use perturbation theory valid in the limit when the induced  world-sheet geometry can be considered as a small deformation of $AdS_2$  govern by the deformation parameter $\theta_0$. This approach has been successfully applied the holographic perturbative computation of a ratio of Wilson loops expectation values  \cite{Forini:2017whz}. Namely, we will expand around the parameter $\alpha = \theta_0^2 $, where the near $AdS_2$ geometry corresponds to the latitude in $S^2 \subset S^5$ parametrized by angle $\theta_0$.  For $\theta_0 = 0$, the worldsheet metric reduces to $AdS_2$.  Under the conditions clarified below we will be able to determine the first leading order correction to the string partition function by the perturbative expansion of the heat kernels. \\

Let $\mathcal{M}$ be a d dimensional smooth compact Riemannian manifold with metric $g_{ij}$ and $\mathcal{O}$ be a second order elliptic operator of the Laplace type. Then, we can define the logarithm of the determinant using $\zeta$-function regularization as, 
\bea
 \log \text{Det}_{\mathcal{M}} \: \mathcal{O} = - \zeta_{\mathcal{O}}^{'} (0),
 \eea
 The $\zeta$ function is related to the integrated heat kernel by the Mellin transform,
\bea
&& \zeta_{\mathcal{O}} (s) = \frac{1}{\Gamma (s)} \: \int_0^{\infty} dt \: t^{s-1} K_{\mathcal{O}} (t), \:\:\:\:\:\:\:\:\:\:\:\:\: K_{\mathcal{O}} (t) = \int d^d x \: \sqrt{g} \: \text{tr} K_{\mathcal{O}} (x,x; t),
\eea
where by construction, $K_{\mathcal{O}} (x, x' ; t)$ satisfies the heat conduction equation
\bea\label{eq1}
&&(\partial_t + \mathcal{O}_x) \: K_{\mathcal{O}} (x, x' ; t) = 0,
\eea
with the initial condition
\bea\label{eq2}
K_{\mathcal{O}} (x, x' ; 0) = \frac{1}{\sqrt{g}} \: \delta^{(d)} \: \big{(} x - x' \big{)} \: \mathbb{I}.
\eea

Let us now assume that  the manifold $\mathcal{M}$ can be viewed as a deformation of another manifold $\bar{\mathcal{M}}$. Namely, for  $\alpha = 0$ we have   $\bar{\mathcal{M}}$ with metric $\bar{g}_{ij}$; we further assume that in this limit the spectral problem can be solved exactly and seek to construct the solution for $\mathcal{M}$. We can expand $K_{\mathcal{O}}$ and subsequently $\text{Det}_{\mathcal{M}} \mathcal{O} $ in perturbation theory in $\alpha$:
\bea\begin{split}\label{Eq:Expansion}
g_{ij} &= \bar{g}_{ij} + \alpha \: \tilde{g}_{ij} + O (\alpha^2), \\
\mathcal{O} &= \bar{\mathcal{O}} + \alpha \: \tilde{\mathcal{O}} + O (\alpha^2), \\
K_{\mathcal{O}} (x, x' ; t) &= \bar{K}_{\mathcal{O}} (x, x' ; t) + \alpha \: \tilde{K}_{\mathcal{O}} (x, x' ; t) + O (\alpha^2),
\end{split}\eea
such that $\bar{K}_{\mathcal{O}} (x, x' ; t)$ satisfies \eqref{eq1} and \eqref{eq2} .  \\

It can be shown that, $\tilde{K}_{\mathcal{O}} (x, x' ; t)$ can be solved from
\bea
&&(\partial_t + \bar{\mathcal{O}}_x) \: \tilde{K}_{\mathcal{O}} (x, x' ; t) + \tilde{\mathcal{O}}_x \: \bar{K}_{\mathcal{O}} (x, x' ; t)  = 0,
\eea
with the initial condition
\bea
\tilde{K}_{\mathcal{O}} (x, x' ; t) = - \frac{\tilde{g}}{2 \: \bar{g}^{3/2}} \: \delta^{(d)} \: (x - x') \: \mathbb{I}.
\eea
The trace of heat kernel can be written as;
\bea\label{eq56}
 \tilde{K}_{\mathcal{O}} (t) &=& - t \: \int d^d x \: \sqrt{\bar{g}} \: \text{tr} \big{[} \tilde{\mathcal{O}}_x \: \bar{K}_{\mathcal{O}} (x, x' ; t)  \big{]}_{x = x'}.
\eea
In perturbation theory, the $\zeta$-function and the determinant takes the form 
\bea
\log \text{Det}_{\mathcal{M}} \mathcal{O} &=& - \bar{\zeta}^{'}_{\mathcal{O}} (0) + - \alpha \: \tilde{\zeta}^{'}_{\mathcal{O}} (0) + O (\alpha^2), \\
\tilde{\zeta}^{'}_{\mathcal{O}} (s) &=& \frac{1}{\Gamma(s)} \: \int_0^{\infty} dt \: t^{s-1} \tilde{K}_{\mathcal{O}} (t).
\eea

In our context, the string partition function corresponding to the Wilson loop in the gauge theory is given by 
\bea
Z = \langle W (\lambda, \alpha) \rangle \equiv e^{- \Gamma}, \:\:\:\:\:\: \Gamma = \sqrt{\lambda} \: \Gamma^{(0)} (\alpha) + \Gamma^{(1)} (\alpha) + O (\lambda^{-1/2})
\eea
where $\Gamma^{(0)} (\alpha)$ is the classical piece and object of current interest is $\Gamma^{(1)} (\alpha)$, which corresponds to the one-loop corrections to the string action. In particular, we are interested in evaluating $\tilde{\Gamma}^{(1)} (0) $.



\subsection{Circular Wilson Loop}

In the limit $\theta_0 = 0$, or $\sigma_0 = \infty$, the operators take the following form;
\bea\begin{split}
&\text{Bosons:} \:\:\:\: \bar{\mathcal{O}}_1 = - \Delta_{\rho, \tau} + 2, \:\:\: \bar{\mathcal{O}}_{2 \pm} = \bar{\mathcal{O}}_{3 \pm} = - \Delta_{\rho, \tau} \\
&\text{Fermions:} \:\:\:\: \bar{\mathcal{O}}_{\alpha,\beta,\gamma} = - \iota \: \slashed{\nabla}_{\rho, \tau}  + \iota\: m \: \sigma_3
\end{split}\eea
where $ 4 \: m = {\alpha + \beta + \gamma - 3 \: \alpha \: \beta \: \gamma} $ with $\alpha, \beta, \gamma = \pm 1$ as follows from the spinor reduction described in appendix  \ref{App:Spinors}.

The integrated $AdS_2$ heat kernel and $\zeta$-function for the massive Laplace operator $- \Delta + m^2$ is known to be,
\bea
\bar{K}_{- \Delta + m^2} (t) &=& \frac{V_{AdS_2}}{2 \:\pi} \: \int_0^{\infty} dv \: v \: \tanh (\pi v) \: e^{-t \: (v^2 + m^2 + \frac{1}{4})} \\
\bar{\zeta}_{-\Delta + m^2} (s) &=&  \frac{V_{AdS_2}}{\pi} \: \bigg{[} \frac{(m^2 + \frac{1}{4})^{1-s} }{2 \: (s-1)} - 2 \: \int_0^{\infty} \: dv \: \frac{v}{(e^{2 \pi v} + 1) \: (v^2 + m^2 + \frac{1}{4})^s} \bigg{]}.
\eea

The regularized determinants for $\theta_0 = 0$ bosonic operators is;
\bea
\bar{\zeta}^{'}_{\mathcal{O}_1} (0) &=& - \frac{25}{12} + \frac{3}{2} \: \log \: 2 \pi - 2 \: \log A, \\
\bar{\zeta}^{'}_{\mathcal{O}_{2 \pm}} (0)  = \bar{\zeta}^{'}_{\mathcal{O}_{3 \pm}} (0) &=& - \frac{1}{12} + \frac{1}{2} \: \log \: 2 \pi - 2 \: \log A, 
\eea
where $A$ is the Glaisher constant. 
The spectrum of the bosonic fluctuations correspond to 2 massive scalars ($m^2 = 2$) and 6 massless scalars.

\bea\begin{split}
\bar{\Gamma}^{(1)}_B (0) &= - \frac{2}{2} \: \bar{\zeta}^{'}_{\mathcal{O}_1} (0) - \frac{6}{2} \: \bar{\zeta}^{'}_{\mathcal{O}_{2 \pm, 3 \pm}} (0) \\
&= \frac{7}{3} - 3 \: \log \: 2 \pi + 8 \: \log A
\end{split}\eea

The standard expression for the $AdS_2$ heat kernel corresponding to the square of the massive Dirac operator $-\slashed{\nabla} + m \: \Gamma^3$ is,
\bea
\bar{K}_{-\slashed{\nabla}^2 + m^2} (t) = \frac{V_{AdS_2}}{\pi} \: \int_0^{\infty} dv \: v \: \coth (\pi v) \: e^{-t \: (v^2 + m^2)}
\eea
and the $\zeta$-function is ,
\bea\begin{split}
\bar{\zeta}_{-\slashed{\nabla}^2 + m^2} (s) =  \frac{V_{AdS_2}}{\pi} \: \bigg{[} \frac{(m^2)^{1-s} }{2 \: (s-1)} + 2 \: \int_0^{\infty} \: dv \: \frac{v}{(e^{2 \pi v} - 1) \: (v^2 + m^2)^s} \bigg{]}.
\end{split}\eea

In the case of fermionic excitations, we have 2 modes with $m^2 = 0$ and 6 modes with $m^2 = 1$.
\bea
\bar{\zeta}'_{m^2 = 0} (0) &=& \frac{1}{3} - 4 \: \log \: A \\
\bar{\zeta}'_{m^2 = 1} (0) &=& - \frac{5}{3} - 4 \: \log \: A + 2 \: \log \: 2 \pi
\eea
Final contribution from fermions is:
\bea\begin{split}
\bar{\Gamma}_F^{(1)} (0) &= - \frac{2}{2} \: \bar{\zeta}'_{m^2 = 0} (0) - \frac{6}{2} \: \bar{\zeta}'_{m^2 = 1} (0)  \\
&= 2 \: \bigg{(} \frac{7}{3} + 8 \: \log \: A - 3 \:  \log \: 2 \pi \bigg{)} 
\end{split}\eea

Thus, the one-loop correction in the circular Wilson loop case is:
\bea\begin{split}
\bar{\Gamma}^{(1)} (0) = \bar{\Gamma}_B^{(1)} (0) - \frac{1}{2} \: \bar{\Gamma}_F^{(1)} (0) = 0 
\end{split}\eea
This result certainly requires further scrutiny\footnote{We acknowledge various discussions in the summer of 2015 with Jewel Ghosh regarding the heat kernel approach to the one-loop effective action of the half BPS configuration.}. Here we simply note that, as it stands, it does not agree with the field theory prediction of \cite{Klemm:2012ii}. It also does not agree with a Gelfand-Yaglom based computation which further involved numerical evaluation \cite{Kim:2012nd}. We leave a proper treatment of the expectation value of the half BPS Wilson loop to a separate work. Here we are mostly concerned with the ratio of expectation values.


\subsection{Difference of one-loop effective actions}

The perturbative expansion of the relevant operators here,
\bea
\mathcal{O}_{i} (\theta_0) &=& \bar{\mathcal{O}}_{i} + \tilde{\mathcal{O}}_{i} \: \theta_0^2 + O (\theta_0^4), \:\:\:\:\:\:\: i = 1, 2 \pm, 3 \pm \\
\mathcal{O}_{\alpha,\beta,\gamma} (\theta_0) &=& \bar{\mathcal{O}}_{\alpha,\beta,\gamma} + \tilde{\mathcal{O}}_{\alpha,\beta,\gamma} \: \theta_0^2 + O (\theta_0^4), \\
\mathcal{O}^2_{\alpha,\beta,\gamma} (\theta_0) &=& \bar{\mathcal{O}}^2_{\alpha,\beta,\gamma} + \theta_0^2 \: \{ \bar{\mathcal{O}}_{\alpha,\beta,\gamma}, \tilde{\mathcal{O}}_{\alpha,\beta,\gamma}\} + O (\theta_0^4).
\eea
where $\{..\}$ denotes the anticommutator of two differential operators. \\
In the expansion scheme of (\ref{Eq:Expansion}), the corresponding perturbative operator is
\bea\begin{split}
\tilde{\mathcal{O}}_1  &= \frac{1}{(1 + \cosh \rho)^2} \: \big{(} \Delta_{\rho, \tau} - 2 \big{)}, \\
 \tilde{\mathcal{O}}_{2 \pm }  &= \frac{1}{ (1 + \cosh \rho)^2} \: \bigg{[} \Delta_{\rho, \tau}  - \frac{1}{2 } \: \big{(}  1 \pm  \iota  \: \partial_{\tau}  \big{)} \bigg{]} \\
\tilde{\mathcal{O}}_{3\pm}  &= \frac{1}{(1 + \cosh \rho)^2} \: \bigg{[} \Delta_{\rho, \tau} - \frac{\sinh^2 \rho}{(1 + \cosh \rho)^2} \: (2 \pm \iota \: \partial_{\tau}) \bigg{]}, 
\end{split}\eea
for the bosonic second order operators. While, for the first order fermionic operator, we have,
\bea\begin{split}
\tilde{\mathcal{O}}_{\alpha, \beta, \gamma} (\theta_0) &=  \frac{1}{2 \: (1 + \cosh \rho)^2} \bigg{[}  \iota \slashed{\nabla} + \frac{\sinh \rho}{1 + \cosh \rho} \big{(} \iota \Gamma^{\underline{0}}\big{)} + \Gamma^{\underline{1}} \: \bigg{(} \frac{\alpha \: (1 - \cosh \rho)^2}{2} - \frac{\beta + \gamma}{4} \: \sinh^2 \rho \bigg{)} \\
& - \frac{ (-1 + 3 \: \beta \: \gamma)}{2} \: \big{(} \iota \Gamma_{\underline{01}}\big{)} + \frac{\alpha (1 - 3 \beta \gamma)}{2} \bigg{]}.
\end{split}\eea


Substituting the $\tilde{O}$ in \eqref{eq56}, we get,

\bea\begin{split}\label{eq5}
\tilde{K}_{\mathcal{O}_1} (t) & = - t \: \int_0^{2 \pi} \: d \tau \: \int_0^{\Lambda} \: d \rho \: \frac{\sinh \rho}{(1 + \cosh \rho)^2} \: \bigg{[}  \big{(} \Delta_{\rho, \tau} - 2 \big{)} \: \bar{K}_{- \Delta + 2} \: (\rho, \tau, \rho', \tau'; t) \bigg{]}_{\rho = \rho' , \tau = \tau'}
\end{split}\eea
We know that $\bar{K}$ satisfies, the following equation;
\bea
(\partial_t - \Delta_{\rho, \tau} + 2) \: \bar{K}_{\mathcal{O}_1} \: (\rho, \tau, \rho', \tau'; t) = 0
\eea
Thus, plugging it back in \eqref{eq5}, we  obtain
\bea
\tilde{K}_{\mathcal{O}_1} (t) = - t \: \int_0^{2 \pi} \: d \tau \: \int_0^{\Lambda} \: d \rho \: \frac{\sinh \rho}{(1 + \cosh \rho)^2} \: \partial_t \: \bar{K}_{\mathcal{O}_1} \: (\rho, \tau, \rho, \tau; t)
\eea
Now we can take the limit $\Lambda \rightarrow \infty$ and using the integral representation of heat kernel $\bar{K}$;
\bea
\tilde{K}_{\mathcal{O}_1} (t) = \frac{t}{2} \:  \int_0^{\infty} \: dv \: v \: \tanh (\pi \: v) \: \bigg{(} v^2 + \frac{9}{4} \bigg{)} \: e^{- t \: \big{(}  v^2  + 9/4 \big{)}}
\eea
Using $\tanh (\pi \: v) = 1 - 2/ (e^{2 \pi v} + 1)$ and we can write the corresponding $\zeta$-function as;
\bea
\tilde{\zeta}_{\mathcal{O}_1} (s) = \int_0^{\infty} \: dv \: \frac{s \: v}{2 \: (v^2 + 9/4)^s}  - \int_0^{\infty} \: dv \: \frac{s \: v}{(e^{2 \pi v } + 1) \: (v^2 + 9/4)^s} 
\eea

The first integral converges only for Re s $>$ 1, we can first integrate over $v$ and then analytically continue to all values of $s$
\bea
\tilde{\zeta}_{\mathcal{O}_1} (s) = \frac{s}{4 \: (s - 1)} \: \bigg{(} \frac{9}{4} \bigg{)}^{1 - s} - s \:\int_0^{\infty} \:  dv \: \frac{ v}{(e^{2 \pi v } + 1) \: (v^2 + 9/4)^s} .
\eea
The final result is
\bea
\tilde{\zeta}^{'}_{\mathcal{O}_1} (0) = - \frac{7}{12}.
\eea
In the case of $\mathcal{O}_{2\pm}$, we will add the contribution from $\mathcal{O}_{2+}$ and $\mathcal{O}_{2- }$ to get rid of the $\partial_{\tau}$ term which simplifies the calculation substantially.
\bea\begin{split}
\tilde{K}_{\mathcal{O}_{2+}} (t) + \tilde{K}_{\mathcal{O}_{2-}} (t) 
&=   t \: \int_0^{\infty} \: d v \: \bigg{[} \bigg{(} v^2  + \frac{3}{4} \bigg{)} \: v \: \tanh (\pi v) \: e^{-t \: \big{(} v^2 + \frac{1}{4} \big{)}} \bigg{]},
\end{split}\eea
\bea\begin{split}
\tilde{\zeta}_{\mathcal{O}_{2+}} (s) + \tilde{\zeta}_{\mathcal{O}_{2-}} (s)
&=    \int_0^{\infty} \: d v \: s v \: \frac{\big{(} v^2  + \frac{3}{4} \big{)}}{\big{(} v^2  + \frac{1}{4} \big{)}^{1 + s}}  - 2 \: s\:  \int_0^{\infty} \: d v \: \frac{ v}{e^{2 \pi v} + 1}  \: \frac{\big{(} v^2  + \frac{3}{4} \big{)}}{\big{(} v^2  + \frac{1}{4} \big{)}^{1 + s}},
\end{split}\eea
\bea
\tilde{\zeta}_{\mathcal{O}_{2+}}^{'} (0) + \tilde{\zeta}_{\mathcal{O}_{2-}}^{'} (0) = - \frac{1}{6} + \frac{\gamma}{2},
\eea
where $\gamma$ is the Euler-Mascheroni constant.

Similarly, for the operator $\mathcal{O}_{3\pm}$, we get
\bea\begin{split}
\tilde{K}_{\mathcal{O}_{3+}} (t) + \tilde{K}_{\mathcal{O}_{3-}} (t)
&=   t \: \int_0^{\infty} \: d v \: \bigg{[} \bigg{(} v^2  + \frac{5}{4} \bigg{)} \: v \: \tanh (\pi v) \: e^{-t \: \big{(} v^2 + \frac{1}{4} \big{)}} \bigg{]},
\end{split}\eea
\bea\begin{split}
\tilde{\zeta}_{\mathcal{O}_{3+}} (s) + \tilde{\zeta}_{\mathcal{O}_{3-}} (s)
&=    s \:\int_0^{\infty} \: d v \: v \: \frac{\big{(} v^2  + \frac{5}{4} \big{)}}{\big{(} v^2  + \frac{1}{4} \big{)}^{1 + s}}  - 2 \: s\:  \int_0^{\infty} \: d v \: \frac{ v}{e^{2 \pi v} + 1}  \: \frac{\big{(} v^2  + \frac{5}{4} \big{)}}{\big{(} v^2  + \frac{1}{4} \big{)}^{1 + s}},
\end{split}\eea
\bea
\tilde{\zeta}_{\mathcal{O}_{3+}}^{'} (0) + \tilde{\zeta}_{\mathcal{O}_{3-}}^{'} (0) = - \frac{1}{6} + \gamma.
\eea

The total contribution for bosonic operators is simply
\bea\begin{split}
\tilde{\Gamma}^{(1)}_B &= - \frac{2}{2} \: \tilde{\zeta}^{'}_{\mathcal{O}_1} (0) - \frac{2}{2} \: \tilde{\zeta}^{'}_{\mathcal{O}_{2+}} (0) - \frac{2}{2} \: \tilde{\zeta}^{'}_{\mathcal{O}_{2-}} (0) - \frac{1}{2} \: \tilde{\zeta}^{'}_{\mathcal{O}_{3+}} (0) - \frac{1}{2} \: \tilde{\zeta}^{'}_{\mathcal{O}_{3-}} (0)  \\
&  = \frac{5}{6} - \gamma.
\end{split}\eea

An important computational ingredient in case of fermions is 
\bea\begin{split}
\{ \bar{\mathcal{O}}_{\alpha,\beta,\gamma}, \tilde{\mathcal{O}}_{\alpha,\beta,\gamma}\} 
&= \{ \bar{\mathcal{O}}, \tilde{\mathcal{O}}\} =  \frac{1}{(1 + \cosh \rho)^2} \slashed{\nabla}^2_{\rho, \tau}  - \frac{m \: (1 - 3 \beta \gamma)}{2 (1 + \cosh \rho)^2}  \\
&+ \frac{\iota X(\rho)}{\sinh^2 \rho (1 + \cosh \rho)^2} \: \partial_{\tau},
\end{split}\eea
where
\bea
X(\rho) =  \frac{\alpha \: (1 - \cosh \rho)^2}{2} - \frac{\beta + \gamma}{4} \: \sinh^2 \rho. \eea

We can derive formal expressions which can be evaluated for the cases of interest, we skip some intermediate steps that involve Mellin transform from the heat kernel to the zeta function.  In particular
\begin{align}
\delta\zeta_F(s)&=\frac{1}{\Gamma(s)}\int_0^\infty dt t^{s-1} \delta K(t)=\int_0^\infty dv \frac{s v \left(v^2+2m^2+m\frac{\alpha(\beta+\gamma)}{2}\right)}{(v^2+m^2)^{s+1}}\coth\pi v\\
&= \int_0^\infty dv \frac{s v (v^2+2m^2)}{(v^2+m^2)^{s+1}} + 2\int_0^\infty dv \frac{s v (v^2+2m^2+m\frac{\alpha(\beta+\gamma)}{2})}{(v^2+m^2)^{s+1}(e^{2\pi v}-1)}\\
&=\frac{m^{1-2s}\left(m(-1+2s)+\frac{\alpha(\beta+\gamma)}{2}(s-1)\right)}{2(s-1)}+ 2\int_0^\infty dv \frac{s v (v^2+2 m^2)}{(v^2+m^2)^{s+1}(e^{2\pi v}-1)},
\end{align}
thus obtaining
\begin{align}
\delta\zeta_F'(0)&=-\frac12 m(m+\left(m+\frac{\alpha(\beta+\gamma)}{2})\ln m^2\right)+2\int_0^\infty dv \frac{v \left(v^2+2 m^2+m\frac{\alpha(\beta+\gamma)}{2}\right)}{(v^2+m^2)(e^{2\pi v}-1)}\\
&=-\frac12 m(m+\left(m+\frac{\alpha(\beta+\gamma)}{2})\ln m^2\right)+2\int_0^\infty dv \frac{v }{(e^{2\pi v}-1)}\\
&\qquad +2m\left(m+\frac{\alpha(\beta+\gamma)}{2}\right)\int_0^\infty dv \frac{v }{(v^2+m^2)(e^{2\pi v}-1)}\\
&=-\frac12 m(m+\left(m+\frac{\alpha(\beta+\gamma)}{2})\ln m^2\right)+\frac{1}{12}+m\left(m+\frac{\alpha(\beta+\gamma)}{2}\right)\left(\frac12\ln m^2-\frac{1}{2|m|}-\psi(|m|)\right),
\end{align}
where $\psi(x)=\frac{d}{dx}\Gamma(x)$ is the digamma function. In particular, evaluating for $m=0$ and $\beta=\gamma$, we obtain
\be
-\alpha=\beta=\gamma \quad \Rightarrow \delta\zeta_F'(0)=-\frac{5}{12},
\ee
while for $m=1$ and $\beta=-\gamma$ leads to
\be
\beta=-\gamma \quad \Rightarrow \delta\zeta_F'(0)=-\frac{11}{12}+\gamma
\ee
and finally for  $m=-1$ and $\beta=\gamma$
\be
\alpha=\beta=\gamma \quad \Rightarrow \delta\zeta_F'(0)=-\frac{5}{12}
\ee
thus adding to the following total fermionic contribution 
\be
\delta\zeta^{\rm tot}_F(s)=\frac12\left[2\times\left(- \frac{5}{12}\right)+4\times \left(-\frac{11}{12}+\gamma\right)+2\times  \left(-\frac{5}{12}\right)\right]=-\frac{8}{3}+2\gamma.
\ee
On the other hand, the bosonic contribution reads
\be
\delta\zeta^{\rm tot}_B(s)=-\frac{5}{3}+2\gamma.
\ee
The total one-loop perturbative contribution is
\bea
\delta\zeta^{\rm tot}_B(0)-\delta\zeta^{\rm tot}_F(0) = \bigg{(} - \frac{5}{3} + 2 \gamma \bigg{)} - \bigg{(} - \frac{8}{3} + 2 \gamma \bigg{)} =1.
\eea

Thus,
\bea
\Delta \Gamma^{\text{1-loop}}_{\text{effective}} (\theta_0) = \frac{1}{2} \theta_0^2 ,
\eea
which agrees with the field theory prediction at the given order.

\section{One-loop Effective Action:  Zeta Function Regularization}\label{Sec:Zeta}
In this section we follow our previous work \cite{Aguilera-Damia:2018rjb,Aguilera-Damia:2018twq} where we developed a regularization in the case of radial determinants that coincides with $\zeta$-function regularization in various cases. There are various reasons to tackle the problem using these methods. First, one would obviously like to go beyond the small $\theta_0$ limit and obtain and expression that is valid in the whole range of $\theta_0$. Second, by construction, our regularization is diffeomorphic invariant and works directly on the disk; other approaches \cite{Forini:2015bgo,Faraggi:2016ekd,Cagnazzo:2017sny} rely on mapping the problem from the disk to the cylinder. Although these latter methods have proven to be quite effective it is conceptually satisfying to deal with the problem directly on the disk.

The main outcome of \cite{Aguilera-Damia:2018rjb} is a prescription for computing $\zeta$-function regularized determinants of radial operators in asymptotically $AdS_2$ spacetimes.  The result for bosons is

\bea\begin{split}
\ln \frac{\text{det} \: \mathcal{O}}{\text{det} \: \mathcal{O}^{\text{free}}} &= \ln \frac{\text{det} \: \mathcal{O}_0}{\text{det} \: \mathcal{O}_0^{\text{free}}} + \sum_{l = 1}^{\infty} \bigg{(} \ln \frac{\text{det} \: \mathcal{O}_l}{\text{det} \: \mathcal{O}_l^{\text{free}}} + \ln \frac{\text{det} \: \mathcal{O}_{-l}}{\text{det} \: \mathcal{O}_{-l}^{\text{free}}} + \frac{2}{l} \: \hat{\zeta}_{\mathcal{O}}(0) \bigg{)} - 2 \: \big{(} \gamma + \ln \frac{\mu}{2} \big{)} \: \hat{\zeta}_{\mathcal{O}}(0) \\
&+ \int_0^{\infty} d \rho \: \sinh \rho \: \ln (\sinh \rho) \: V - q^2 \: \int_0^{\infty} d \rho \: \frac{\mathcal{A}^2}{\sinh \rho},
\end{split}\eea
\bea\begin{split}
\hat{\zeta}_{\mathcal{O}}(0) = - \frac{1}{2} \int_0^{\infty} d\rho \sinh \rho \: V,
\end{split}\eea
whereas for fermions, it reads
\bea\begin{split}
\ln \frac{\text{det} \: \mathcal{O}}{\text{det} \: \mathcal{O}^{\text{free}}} &=  \sum_{l = \frac{1}{2}}^{\infty} \bigg{(} \ln \frac{\text{det} \: \mathcal{O}_l}{\text{det} \: \mathcal{O}_l^{\text{free}}} + \ln \frac{\text{det} \: \mathcal{O}_{-l}}{\text{det} \: \mathcal{O}_{-l}^{\text{free}}} + \frac{2}{l + \frac{1}{2}} \: \hat{\zeta}_{\mathcal{O}}(0) \bigg{)} - 2 \: \big{(} \gamma + \ln \frac{\mu}{2} \big{)} \: \hat{\zeta}_{\mathcal{O}}(0) \\
&+ \int_0^{\infty} d \rho \: \sinh \rho \: \ln (\sinh \rho) \: \big{(} (m + V)^2 - W^2 - m^2 \big{)} - q^2 \: \int_0^{\infty} d \rho \: \frac{\mathcal{A}^2}{\sinh \rho} - \int_0^{\infty} d \rho \: \sinh \rho \: W^2,
\end{split}\eea
\bea\begin{split}
\hat{\zeta}_{\mathcal{O}}(0) = - \frac{1}{2} \int_0^{\infty} d\rho \sinh \rho \: \big{(} (m + V)^2 - W^2 - m^2 \big{)},
\end{split}\eea

\subsection{Bosons}
We now proceed to apply the prescription above to the different bosonic operators. 
\subsubsection{$\big{(} \chi^{2,3 }\big{)} \: \: \mathcal{O}_1 (\theta_0)$}

The action for these fluctuations is

\bea\begin{split}
\mathcal{O}_1 (\theta_0) &= M^{-1} \big{(} -g^{\mu \nu} \: \nabla_{\mu} \nabla_{\nu} + 2 \big{)}
\end{split}\eea
We see that the rescaled operator does not depend on $\theta_0$, meaning that these fluctuations contribute only with an anomaly.

\bea\begin{split}
\ln \bigg{(} \frac{\text{det}\: \mathcal{O}_1 (\theta_0)}{\text{det}\: \mathcal{O}_1 (0)}\bigg{)} &= \theta_0  \sin \theta_0 + \frac{1}{2} \: \sin^2 \frac{\theta_0}{2} + \bigg{(} \frac{7}{3} + 2 \: \cos \theta_0 \bigg{)} \: \ln \cos \frac{\theta_0}{2} \\
&= \frac{7}{12} \: \theta_0^2 + O \big{(} \theta_0^4 \big{)}
\end{split}\eea

\subsubsection{$\big{(} \chi^{4,5 }\big{)} \: \: \mathcal{O}_{2 \pm} (\theta_0)$}

For charged fluctuations, we have:

\begin{empheq}{alignat=7}
	\ln\left(\frac{\det\mathcal{O}_{AdS_2}(\theta_0)}{\det\mathcal{O}_{AdS_2}(0)}\right)&=\ln\frac{\psi_0(\theta_0)}{\psi_0(0)}+\sum_{l=1}^{\infty}\left(\ln\frac{\psi_l(\theta_0)}{\psi_l(0)}+\ln\frac{\psi_{-l}(\theta_0)}{\psi_{-l}(0)}-\frac{D}{l}\right)+F+D\gamma\,,
\end{empheq}
where 
\begin{empheq}{alignat=7}
	D&\equiv\int_0^{\infty}d\rho\,\sinh\rho\,V_{AdS_2}(\rho)\,
	\\
	F&\equiv\int_0^{\infty}d\rho\left(\sinh\rho\,V_{AdS_2}(\rho)\ln\left(\frac{\sinh\rho}{2}\right)-\frac{A(\rho)^2}{\sinh\rho}\right)\,,
\end{empheq}

The relevant operator in $AdS_2$ is
\begin{empheq}{alignat=7}
	\mathcal{O}_{AdS_2}&=-\frac{1}{\sinh\rho}\partial_{\rho}\left(\sinh\rho\,\partial_{\rho}\right)+\frac{\left(l-A(\rho)\right)^2}{\sinh^2\rho}+V_{AdS_2}\,,
\end{empheq}
where the gauge field and the potential read
\begin{empheq}{alignat=7}
	A(\rho)&=-\frac{\left(\cosh\rho-1\right)^2\left(1-\cos\theta_0\right)}{\cosh^2\rho+2\cosh\rho\cos\theta_0+1}\,,
	&\qquad
	V_{AdS_2}(\rho)&=\frac{\partial_{\rho}A(\rho)}{\sinh\rho}\,.
\end{empheq}
Notice that we can write this as
\begin{empheq}{alignat=7}
	A(\rho)&=\sinh\rho\,\partial_{\rho}W(\rho)\,,
	&\qquad
	W(\rho)&=\frac{1}{2}\ln\left(\frac{\left(\cosh\rho+1\right)^2}{\cosh^2\rho+2\cosh\rho\cos\theta_0+1}\right)\,.
\end{empheq}
This fact allows us to write the solution to the equation of motion as
\begin{empheq}{alignat=7}
	f_l(\rho)&=\tanh^{-l}\left(\frac{\rho}{2}\right)e^{W(\rho)}\left(A+B\int d\rho\frac{\tanh^{2l}\left(\frac{\rho}{2}\right)e^{-2W(\rho)}}{\sinh\rho}\right)\,.
\end{empheq}
For the case at hand, the regular solution at $\rho=0$ is
\begin{empheq}{alignat=7}
	f_l(\rho)&=\left\{
	\begin{array}{cc}
		{\displaystyle2^{-\left(l+\frac{1}{2}\right)}\sqrt{1+\cos\theta_0}\tanh^{-l}\left(\frac{\rho}{2}\right)\frac{\cosh\rho+1}{\sqrt{\cosh^2\rho+2\cosh\rho\cos\theta_0+1}}} & l<0
		\\\\
		{\displaystyle\frac{2^{l+\frac{1}{2}}\tanh^{l}\left(\frac{\rho}{2}\right)}{\left(l+2\right)\sqrt{1+\cos\theta_0}}\frac{\sqrt{\cosh^2\rho+2\cosh\rho\cos\theta_0+1}}{\cosh\rho+1}\left(l+\frac{\left(\cosh\rho+1\right)^2\left(1+\cos\theta_0\right)}{\cosh^2\rho+2\cosh\rho\cos\theta_0+1}\right)} & l>0
	\end{array}
	\right.\,.
\end{empheq}
We then find
\begin{empheq}{alignat=7}
	\psi_l(\theta_0)&=\left\{
	\begin{array}{cc}
		{\displaystyle\left(\frac{1+\cos\theta_0}{2}\right)^{\frac{1}{2}}} & l\leq0
		\\\\
		{\displaystyle\left(\frac{1+\cos\theta_0}{2}\right)^{-\frac{1}{2}}\left(\frac{l+1+\cos\theta_0}{l+2}\right)} & l\geq0
	\end{array}
	\right.\,.
\end{empheq}
Next, we compute the integrals
\bea\begin{split}
	D&\equiv\int_0^{\infty}d\rho\,\sinh\rho\,V_{AdS_2}(\rho)
	\\
	&=-2\sin^2\frac{\theta_0}{2}\,,
	\\
	F&\equiv\int_0^{\infty}d\rho\left(\sinh\rho\,V_{AdS_2}(\rho)\ln\left(\frac{\sinh\rho}{2}\right)-\frac{A(\rho)^2}{\sinh\rho}\right)\,,
	\\
	&=-\frac{\theta_0}{2}\sin\theta_0+\left(2+\cos\theta_0\right)\ln\cos\frac{\theta_0}{2}+\sin^2\frac{\theta_0}{2}\,,
	\\
	I&\equiv\frac{1}{2}\sin^2\frac{\theta_0}{2}+\frac{1}{3}\ln\cos\frac{\theta_0}{2}+\frac{1}{2}\int d\rho\,\sinh\rho\ln M\,V_{AdS_2}
	\\
	&=-\frac{1}{2}\sin^2\frac{\theta_0}{2}+\frac{1}{3}\ln\cos\frac{\theta_0}{2}+\frac{1}{2}\theta_0\sin\theta_0+2\cos^2\frac{\theta_0}{2}\ln\cos\frac{\theta_0}{2}\,.
\end{split}\eea
Finally, putting everything together we get
\bea\begin{split}
	\Omega_{\textrm{ind}}(\theta_0)&=\ln\frac{\psi_0(\theta_0)}{\psi_0(0)}+\sum_{l=1}^{\infty}\left(\ln\frac{\psi_l(\theta_0)}{\psi_l(0)}+\ln\frac{\psi_{-l}(\theta_0)}{\psi_{-l}(0)}-\frac{D}{l}\right)+F+D\gamma
	\\
	&=\ln\cos\frac{\theta_0}{2}-\ln\Gamma\left(2\cos^2\frac{\theta_0}{2}\right)-2\ln\cos\frac{\theta_0}{2}+2\gamma\sin^2\frac{\theta_0}{2}
	\\
	&-\frac{\theta_0}{2}\sin\theta_0+\left(2+\cos\theta_0\right)\ln\cos\frac{\theta_0}{2}+\sin^2\frac{\theta_0}{2}-2\gamma\sin^2\frac{\theta_0}{2}
	\\
	&-\frac{1}{2}\sin^2\frac{\theta_0}{2}+\frac{1}{3}\ln\cos\frac{\theta_0}{2}+\frac{1}{2}\theta_0\sin\theta_0+2\cos^2\frac{\theta_0}{2}\ln\cos\frac{\theta_0}{2}
	\\
	&=-\ln\Gamma\left(\cos\theta_0\right)-\ln\cos\theta_0+\left(\frac{7}{3}+2\cos\theta_0\right)\ln\cos\frac{\theta_0}{2}+\frac{1}{2}\sin^2\frac{\theta_0}{2}\,.
\end{split}\eea
As before, the small $\theta_0$ expansion coincides with the results of \cite{Forini:2017whz}:
\begin{empheq}{alignat=7}
	\Omega_{\textrm{ind}}(\theta_0)&=\frac{1}{2}\left(\frac{1}{6}-\gamma\right)\theta_0^2+O\left(\theta_0^4\right)\,.
\end{empheq}

\subsubsection{$\big{(} \chi^{6,7,8,9 }\big{)} \: \: \mathcal{O}_{3 \pm} (\theta_0)$}

The relevant operator in $AdS_2$ is
\begin{empheq}{alignat=7}
	\mathcal{O}_{AdS_2}&=-\frac{1}{\sinh\rho}\partial_{\rho}\left(\sinh\rho\,\partial_{\rho}\right)+\frac{\left(l-B(\rho)\right)^2}{\sinh^2\rho}+V_{AdS_2}\,,
\end{empheq}
where the gauge field and the potential read
\begin{empheq}{alignat=7}
	B(\rho)&=\frac{1}{2}\frac{\left(\cosh\rho-1\right)\left(1-\cos\theta_0\right)}{\cosh\rho+\cos\theta_0}\,,
	&\qquad
	V_{AdS_2}(\rho)&=-\frac{\partial_{\rho}B(\rho)}{\sinh\rho}\,.
\end{empheq}
Notice that we can write this as
\begin{empheq}{alignat=7}
	B(\rho)&=\sinh\rho\,\partial_{\rho}W(\rho)\,,
	&\qquad
	W(\rho)&=\frac{1}{2}\ln\left(\frac{\left(\cosh\rho-1\right)\left(\cosh\rho+\cos\theta_0\right)}{\sinh^2\rho}\right)\,.
\end{empheq}
This fact allows us to write the solution to the equation of motion as
\begin{empheq}{alignat=7}
	f_l(\rho)&=\tanh^{l}\left(\frac{\rho}{2}\right)e^{-W(\rho)}\left(A+B\int d\rho\frac{\tanh^{-2l}\left(\frac{\rho}{2}\right)e^{2W(\rho)}}{\sinh\rho}\right)\,.
\end{empheq}
For the case at hand, the regular solution at $\rho=0$ is
\begin{empheq}{alignat=7}
	f_l(\rho)&=\left\{
	\begin{array}{cc}
		{\displaystyle2^l\cos\frac{\theta_0}{2}\tanh^{l-\frac{1}{2}}\left(\frac{\rho}{2}\right)\sqrt{\frac{\sinh\rho}{\cosh\rho+\cos\theta_0}}} & l>0
		\\\\
		{\displaystyle\frac{\tanh^{-l+\frac{1}{2}}\left(\frac{\rho}{2}\right)}{2^{l+1}\left(l-1\right)\cos\frac{\theta_0}{2}}\sqrt{\frac{\cosh\rho+\cos\theta_0}{\sinh\rho}}\left(2l-\frac{\left(\cosh\rho-1\right)\left(1+\cos\theta_0\right)}{\cosh\rho+\cos\theta_0}\right)} & l<0
	\end{array}
	\right.\,.
\end{empheq}
We then find
\begin{empheq}{alignat=7}
	\psi_l(\theta_0)&=\left\{
	\begin{array}{cc}
		{\displaystyle\left(\frac{1+\cos\theta_0}{2}\right)^{\frac{1}{2}}} & l\leq0
		\\\\
		{\displaystyle\left(\frac{1+\cos\theta_0}{2}\right)^{-\frac{1}{2}}\left(\frac{l-\frac{1+\cos\theta_0}{2}}{l-1}\right)} & l\leq0
	\end{array}
	\right.\,.
\end{empheq}
Next, we compute the integrals
\bea\begin{split}
	D&\equiv\int_0^{\infty}d\rho\,\sinh\rho\,V_{AdS_2}(\rho)
	\\
	&=-\sin^2\frac{\theta_0}{2}\,,
	\\
	F&\equiv\int_0^{\infty}d\rho\left(\sinh\rho\,V_{AdS_2}(\rho)\ln\left(\frac{\sinh\rho}{2}\right)-\frac{A(\rho)^2}{\sinh\rho}\right)
	\\
	&=2\cos^2\frac{\theta_0}{2}\ln\cos\frac{\theta_0}{2}+\frac{1}{2}\sin^2\frac{\theta_0}{2}\,,
	\\
	I&\equiv\frac{1}{2}\sin^2\frac{\theta_0}{2}+\frac{1}{3}\ln\cos\frac{\theta_0}{2}+\frac{1}{2}\int d\rho\,\sinh\rho\ln M\,V_{AdS_2}
	\\
	&=\frac{3}{2}\sin^2\frac{\theta_0}{2}+\frac{1}{3}\ln\cos\frac{\theta_0}{2}-\frac{1}{4}\theta_0\sin\theta_0+\sin^2\frac{\theta_0}{2}\ln\cos\frac{\theta_0}{2}\,.
\end{split}\eea
Finally, putting everything together we get
\bea\begin{split}
	\Omega_{\textrm{ind}}(\theta_0)&=\ln\frac{\psi_0(\theta_0)}{\psi_0(0)}+\sum_{l=1}^{\infty}\left(\ln\frac{\psi_l(\theta_0)}{\psi_l(0)}+\ln\frac{\psi_{-l}(\theta_0)}{\psi_{-l}(0)}-\frac{D}{l}\right)+F+D\gamma
	\\
	&=\ln\cos\frac{\theta_0}{2}-\ln\Gamma\left(\cos^2\frac{\theta_0}{2}\right)-2\ln\cos\frac{\theta_0}{2}+\gamma\sin^2\frac{\theta_0}{2}
	\\
	&+2\cos^2\frac{\theta_0}{2}\ln\cos\frac{\theta_0}{2}+\frac{1}{2}\sin^2\frac{\theta_0}{2}-\gamma\sin^2\frac{\theta_0}{2}
	\\
	&+\frac{3}{2}\sin^2\frac{\theta_0}{2}+\frac{1}{3}\ln\cos\frac{\theta_0}{2}-\frac{1}{4}\theta_0\sin\theta_0+\sin^2\frac{\theta_0}{2}\ln\cos\frac{\theta_0}{2}
	\\
	&=-\ln\Gamma\left(\cos^2\frac{\theta_0}{2}\right)+\frac{1}{2}\left(\frac{5}{3}+\cos\theta_0\right)\ln\cos\frac{\theta_0}{2}-\frac{1}{4}\theta_0\sin\theta_0+2\sin^2\frac{\theta_0}{2}\,.
\end{split}\eea
The small $\theta_0$ expansion is
\begin{empheq}{alignat=7}
	\Omega_{\textrm{ind}}(\theta_0)&=\frac{1}{2}\left(\frac{1}{6}-\frac{\gamma}{2}\right)\theta_0^2+O\left(\theta_0^4\right)\,.
\end{empheq}
This coincides with the perturbative heat kernel approach. \\

The total bosonic contribution is: 

\bea\begin{split}
&\Rightarrow \frac{1}{2} \: \bigg{[} 2 \: \ln \bigg{(} \frac{\text{det}\: \mathcal{O}_1 (\theta_0)}{\text{det}\: \mathcal{O}_1 (0)} \bigg{)} + 2 \: \ln \bigg{(} \frac{\text{det}\: \mathcal{O}_{2\pm} (\theta_0)}{\text{det}\: \mathcal{O}_{2\pm} (0)}\bigg{)} + 4 \: \ln \bigg{(} \frac{\text{det}\: \mathcal{O}_{3\pm} (\theta_0)}{\text{det}\: \mathcal{O}_{3\pm} (0)}\bigg{)}\bigg{]}  =
\end{split}\eea

\bea\begin{split}
&= \frac{\theta_0}{2}  \sin \theta_0 + 5 \: \sin^2 \frac{\theta_0}{2} + \bigg{(} \frac{19}{3} + 5 \: \cos \theta_0 \bigg{)} \: \ln \cos \frac{\theta_0}{2}  - 2 \: \ln \: \Gamma \bigg{(} \cos^2 \frac{\theta_0}{2} \bigg{)} - \ln \big{(} \Gamma (\cos \theta_0)\big{)} - \ln \big{(} \cos \theta_0\big{)}  \\
&= \bigg{(} \frac{5}{6} - \gamma \bigg{)} \: \theta_0^2 + O (\theta_0^4)
\end{split}\eea
which matches with the perturbative heat kernel calculation.


\subsection{Fermions}


\subsubsection{$\beta = - \gamma$  }

Then, following quantities simplifies to:
\bea
D_{\mu} = \nabla_{\mu} + \iota \frac{\alpha}{2} \: \mathcal{A}_{\mu}, \:\:\:\: V (\rho) = \frac{1}{\sqrt{M (\rho)}} - 1, \:\:\:\: W (\rho) = \frac{\sin^2 \theta (\rho)}{\sqrt{M (\rho)} \: \sinh^2 \rho}.
\eea
Take $\Gamma^{\underline{0}} = \sigma_1$ and $ \Gamma^{\underline{1}} = \sigma_2$. Consider operator of form,
\bea
\mathcal{O}_{\alpha} (\theta_0) =  - \iota \slashed{D} + V_1
\eea
where
\bea
V_1 = - \frac{\partial_{\rho} M}{4 \: M} \: \iota \sigma_1  +  \frac{1}{\sqrt{M}} \bigg{(} \sigma_3  + \alpha \frac{\sin^2 \theta (\rho)}{ \sinh^2 \rho} \bigg{)}.
\eea
Using circular symmetry, we can expand this into Fourier components. Explicitly,
\bea
 \iota \:\mathcal{O}_l = 
\begin{pmatrix} 
 \frac{\iota}{\sqrt{M}} \Big{(} 1 + \alpha \frac{\sin^2 \theta (\rho)}{ \sinh^2 \rho} \Big{)} & \partial_{\rho} + \frac{\coth \rho}{2} + \frac{\partial_{\rho} M}{4 \: M} - \frac{l}{\sinh \rho} - \frac{\alpha \: \mathcal{A}}{2 \sinh \rho} \\
 \partial_{\rho} + \frac{\coth \rho}{2} + \frac{\partial_{\rho} M}{4 \: M} + \frac{l}{\sinh \rho} + \frac{\alpha \: \mathcal{A}}{2 \sinh \rho} & \frac{\iota}{\sqrt{M}} \Big{(} - 1 + \alpha \frac{\sin^2 \theta (\rho)}{ \sinh^2 \rho} \Big{)}
\end{pmatrix}\eea

For $\alpha = 1$,
\bea
\ln \bigg{(} \frac{\text{det} \: \mathcal{O}_M(\theta_0) }{\text{det} \mathcal{O}_M (0)} \bigg{)} = \frac{\theta_0}{2} \: \sin \theta_0 + \bigg{(} \frac{7}{3} + 2 \: \cos \theta_0 \bigg{)} \: \ln \cos \frac{\theta_0}{2} - \ln \Gamma (\cos \theta_0 ) - \ln \cos \theta_0
\eea
The relevant integrals in this case are:
\bea
\hat{\zeta}_{\mathcal{O}} (0) = - \frac{1}{2} \int_0^{\infty} d\rho \: \sinh \rho \: \big{(}(m + V)^2 - m^2 - W^2 \big{)} =  \sin^2 \frac{\theta_0}{2}
\eea
\bea
\int_0^{\infty} d\rho \: \sinh \rho \: \ln \bigg{(} \frac{\sinh \rho}{2}\bigg{)} \big{(}(m + V)^2 - m^2 - W^2 \big{)} = 2 \cos \theta_0 \: \ln \cos \frac{\theta_0}{2}
\eea

\bea
\int_0^{\infty} d\rho \: \sinh \rho \: W^2 =  - \frac{1}{2}  \theta_0 \: \sin \theta_0 + 2 \: \sin^2 \frac{\theta_0}{2} 
\eea

\bea
\int_0^{\infty} d\rho \frac{\mathcal{A}^2}{\sinh \rho} = - \sin^2 \frac{\theta_0}{2} -2 \: \log \cos \frac{\theta_0}{2}
\eea
\bea
\int_0^{\infty} d\rho \frac{\mathcal{B}^2}{\sinh \rho} = - \frac{1}{2} \sin^2 \frac{\theta_0}{2} - \log \cos \frac{\theta_0}{2}
\eea

The Weyl anomaly contribution in this case,
\bea\begin{split}
&\frac{1}{4\pi} \: \int d^2 \sigma \sqrt{g} \ln M \: \bigg{[} (m + V)^2 - W^2 + \frac{1}{12} R - \frac{1}{24} \nabla^2 \ln M \bigg{]} \\
&= \frac{1}{4\pi} \: \int d^2 \sigma \sqrt{g} \ln M \: \bigg{[} 2 - M + \frac{1}{12} R - \frac{1}{24} \nabla^2 \ln M \bigg{]} \\
&= \frac{7}{4} \sin^2 \frac{\theta_0}{2} + \frac{11}{6} \ln \cos \frac{\theta_0}{2}
\end{split} \eea


\subsubsection{$\alpha = \beta = \gamma$ }
In this case, 
\bea
D_{\mu} = \nabla_{\mu} + \iota \alpha \: \bigg{(}  \frac{ \mathcal{A}_{\mu}}{2} + \mathcal{B}_{\mu} \bigg{)}, \:\:\:\: V (\rho) = - \frac{1}{2 \: \sqrt{M}} -  \frac{1}{2} \: \sqrt{M} + 1, \:\:\:\: W (\rho) = - \frac{1}{2} \: \frac{\sin^2 \theta (\rho)}{\sqrt{M} \: \sinh^2 \rho}.
\eea
The radial problem becomes, $\mathcal{O}_l \psi_l = 0$,
\bea
 \mathcal{O}_l = - \iota \sigma_1\bigg{(} \partial_{\rho} + \frac{1}{2} \coth \rho + \frac{1}{4} \partial_{\rho}  \ln M \bigg{)} - \frac{1}{\sinh \rho}\sigma_2 \bigg{(} l + \frac{\alpha}{2} \mathcal{A} + \alpha \mathcal{B}\bigg{)} + \sigma_3 (- 1 + V) + \alpha W,
\eea
with $l \in \mathbb{Z} + \frac{1}{2}$. Let
\bea
\psi_l (\rho) = \begin{bmatrix}
u_l (\rho) \\
v_l (\rho)
\end{bmatrix}
\eea

Using circular symmetry, we can expand this into Fourier components. Explicitly,
\bea
 \iota \:\mathcal{O}^{\alpha}_l = 
\begin{pmatrix} 
 \frac{\iota}{2 \: \sqrt{M}} \Big{(} - 1 - M  - \alpha \frac{\sin^2 \theta (\rho)}{ \sinh^2 \rho} \Big{)} & \partial_{\rho} + \frac{\coth \rho}{2} + \frac{\partial_{\rho} M}{4 \: M} - \frac{l}{\sinh \rho} - \frac{\alpha \: (\mathcal{A} + 2 \: \mathcal{B})}{2 \:\sinh \rho} \\
 \partial_{\rho} + \frac{\coth \rho}{2} + \frac{\partial_{\rho} M}{4 \: M} + \frac{l}{\sinh \rho} + \frac{\alpha \: (\mathcal{A} + 2 \: \mathcal{B})}{2 \sinh \rho} & \frac{\iota}{2 \: \sqrt{M}} \Big{(}  1 + M - \alpha \frac{\sin^2 \theta (\rho)}{ \sinh^2 \rho} \Big{)}
\end{pmatrix}\eea

Take $\alpha = 1$,
\bea
 \iota \:\mathcal{O}_l = 
\begin{pmatrix} 
- \iota \: \sqrt{M} & \partial_{\rho} + \frac{\coth \rho}{2} + \frac{\partial_{\rho} M}{4 \: M} - \frac{l}{\sinh \rho} - \frac{ (\mathcal{A} + 2 \: \mathcal{B})}{2 \:\sinh \rho} \\
 \partial_{\rho} + \frac{\coth \rho}{2} + \frac{\partial_{\rho} M}{4 \: M} + \frac{l}{\sinh \rho} + \frac{ (\mathcal{A} + 2 \: \mathcal{B})}{2 \sinh \rho} & \frac{\iota}{ \sqrt{M}} \end{pmatrix}\eea

Take
\bea
\psi_l (\rho) = \begin{bmatrix}
u_l (\rho) \\
v_l (\rho)
\end{bmatrix}
\eea
Now, the system of equations become:
\bea
\bigg{(} \partial_{\rho} + \frac{\coth \rho}{2} + \frac{\partial_{\rho} M}{4 \: M} - \frac{l}{\sinh \rho} - \frac{ (\mathcal{A} + 2 \: \mathcal{B})}{2 \:\sinh \rho} \bigg{)} \: v_l (\rho) - \iota \sqrt{M} \: u_l ( \rho) = 0,  \\
\bigg{(} \partial_{\rho} + \frac{\coth \rho}{2} + \frac{\partial_{\rho} M}{4 \: M} + \frac{l}{\sinh \rho} + \frac{ (\mathcal{A} + 2 \: \mathcal{B})}{2 \:\sinh \rho} \bigg{)} \: u_l (\rho) + \iota \: \frac{1}{\sqrt{M}} \: v_l (\rho) = 0.
\eea
Let 
\bea
D^{\pm} = \partial_{\rho} + \frac{\coth \rho}{2} + \frac{\partial_{\rho} M}{4 \: M} \pm \bigg{(} \frac{l}{\sinh \rho} + \frac{ (\mathcal{A} + 2 \: \mathcal{B})}{2 \:\sinh \rho} \bigg{)}
\eea
We will first solve the second order equation for $v_l (\rho)$. It takes the form,
\bea
 \sqrt{M} \: D^{+} \bigg{(} \frac{1}{\sqrt{M}} \: D^{-} v_l \bigg{)} - v_l = 0.
\eea
We can rewrite the equation as,
\bea
 - \frac{1}{\sinh \rho} \partial_{\rho} \big{(} \sinh \rho \:\partial_{\rho} \:v_l (\rho )\big{)} + \frac{(l + \mathcal{X})^2}{\sinh^2 \rho} v_l (\rho) - \frac{\partial_{\rho} \mathcal{X}}{\sinh \rho} v_l (\rho)  = 0 ,
\eea

where
\bea
\mathcal{X} = \sinh \rho \: \bigg{(}  - \frac{\coth \rho}{2} - \frac{\partial_{\rho} M}{4 \: M}  \bigg{)} +  \frac{\mathcal{A} + 2 \mathcal{B} }{2}.
\eea
\bea
v_l (\rho) = \bigg{(} \tanh \frac{\rho}{2}\bigg{)}^{-l + \frac{1}{2}} \: e^{- \mathcal{W} (\rho)} \: \bigg{(} C_1 + C_2 \: \int d \rho \bigg{(} \tanh \frac{\rho}{2}\bigg{)}^{2 l -1} \frac{e^{2 \mathcal{W} (\rho) }}{\sinh \rho}, \:\:\:\: \partial_{\rho} \mathcal{W} (\rho) = \frac{\mathcal{X} (\rho) + \frac{1}{2}}{\sinh \rho} \bigg{)}.
\eea
Since $\mathcal{W} (\rho)$ is finite at $\rho = 0$. We fix constants $C_1$ and $C_2$ by demanding solution being regular at origin ($\rho = 0$). \\

For $l \geq 1/2 $,
\bea
v_l^{+} (\rho) = C_2 \: \frac{(2 l + \cosh \rho)}{(4 l^2 - 1) \: \sinh \frac{\rho}{2}} \bigg{(} \tanh \frac{\rho}{2}\bigg{)}^{l + \frac{1}{2}} 
\eea 
\bea
u_l^{+} (\rho) = - C_2 \: \frac{  2 \iota\: \sinh \frac{\rho}{2} \: (\cos \theta_0 + \cosh \rho) }{(4 l^2 - 1) \: \sqrt{1 + \cosh^2 \rho + 2  \cos \theta_0 \: \cosh \rho}} \: \bigg{(} \tanh \frac{\rho}{2}\bigg{)}^{l - \frac{1}{2}}
\eea
and for $l \leq -1/2 $,
\bea
v_l^{-} (\rho) =  C_1 \: \cosh \frac{\rho}{2} \: \bigg{(} \tanh \frac{\rho}{2}\bigg{)}^{-l + \frac{1}{2}}
\eea
\bea
u_l^{-} (\rho) = C_1 \: \frac{  \iota\: ( 2 l - \cosh \rho) \: (\cos \theta_0 + \cosh \rho)}{2 \: \cosh \frac{\rho}{2} \: \sqrt{1 + \cosh^2 \rho + 2  \cos \theta_0 \: \cosh \rho}} \: \bigg{(} \tanh \frac{\rho}{2}\bigg{)}^{-l - \frac{1}{2}}
\eea

Thus, for this case we need to evaluate
\bea
\overset{\infty}{\underset{l = \frac{1}{2}}{\sum}} \bigg{(} \ln \frac{\text{det} \:\mathcal{O}_l}{\text{det}^{\text{free}} \mathcal{O}_l} + \ln \frac{\text{det}\: \mathcal{O}_{-l}}{\text{det}^{\text{free}} \mathcal{O}_{-l}}   + \frac{2}{l + \frac{1}{2}} \: \hat{\zeta}_{\mathcal{O}} (0) \bigg{)}
\eea

The relevant formulas in this case, corresponding to $m = -1$, are
\bea
 \big{(} (m + V)^2 - m^2 - W^2 \big{)} = 0 \eea
\bea
\hat{\zeta}_{\mathcal{O}} (0) &=& - \frac{1}{2} \int_0^{\infty} d\rho \: \sinh \rho \: \big{(} (m + V)^2 - m^2 - W^2 \big{)} = 0
\eea
and
\bea
 \int_0^{\infty} d\rho \: \sinh \rho \: \ln \bigg{(} \frac{\sinh \rho}{2}\bigg{)}\: \big{(} (m + V)^2 - m^2 - W^2 \big{)} = 0
\eea
together with
\bea
\int_0^{\infty} d\rho \: \sinh \rho \: W^2 =  - \frac{1}{8}  \theta_0 \: \sin \theta_0 + \frac{1}{2} \: \sin^2 \frac{\theta_0}{2} 
\eea

The Weyl anomaly contribution in this case,
\bea\begin{split}
&\frac{1}{4\pi} \: \int d^2 \sigma \sqrt{g} \ln M \: \bigg{[} (m + V)^2 - W^2 + \frac{1}{12} R - \frac{1}{24} \nabla^2 \ln M \bigg{]} \\
&= \frac{1}{4\pi} \: \int d^2 \sigma \sqrt{g} \ln M \: \bigg{[} 1 + \frac{1}{12} R - \frac{1}{24} \nabla^2 \ln M \bigg{]} \\
&= \frac{\theta_0}{2} \sin \theta_0 +2 \cos^2 \frac{\theta}{2} \: \log \cos \frac{\theta_0}{2} - \frac{1}{4} \sin^2 \frac{\theta_0}{2} - \frac{1}{6} \log \cos \frac{\theta_0}{2}
\end{split}\eea

\bea\begin{split}
\ln \bigg{(} \frac{\text{det} \:O_{\pm}}{\text{det} \:O_{\pm}^{\text{free}}}\bigg{)} = - q^2 \int_0^{\infty} d \rho \: \frac{(\mathcal{A} + 2 \mathcal{B})^2}{\sinh \rho} - \int_0^{\infty} d \rho \sinh \rho \: W^2.
\end{split}\eea

\subsubsection{$ \beta = \gamma = - \alpha$ }

In this case,
\bea
D_{\mu} = \nabla_{\mu} + \iota \alpha \: \bigg{(}  \frac{ \mathcal{A}_{\mu}}{2} - \mathcal{B}_{\mu} \bigg{)}, \:\:\:\: V (\rho) = - \frac{1}{2 \: \sqrt{M}} + \frac{1}{2} \: \sqrt{M} , \:\:\:\: W (\rho) = - \frac{1}{2} \: \frac{\sin^2 \theta (\rho)}{\sqrt{M} \: \sinh^2 \rho}.
\eea
Consider an operator of form,
\bea
\mathcal{O}_{\alpha} (\theta_0) =  - \iota \slashed{D} + V,
\eea
where
\bea
V = - \frac{\partial_{\rho} M}{4 \: M} \: \iota \sigma_1  +  \frac{1}{2 \:\sqrt{M}} \bigg{(} \big{(} - 1 + M \big{)} \sigma_3  - \alpha \frac{\sin^2 \theta (\rho)}{ \sinh^2 \rho} \bigg{)}.
\eea
Using circular symmetry, we can expand this into Fourier components. Explicitly,

\bea
 \iota \:\mathcal{O}^{\alpha}_l = 
\begin{pmatrix} 
 \frac{\iota}{2 \: \sqrt{M}} \Big{(} - 1 +  M  - \alpha \frac{\sin^2 \theta (\rho)}{ \sinh^2 \rho} \Big{)} & \partial_{\rho} + \frac{\coth \rho}{2} + \frac{\partial_{\rho} M}{4 \: M} - \frac{l}{\sinh \rho} - \frac{\alpha \: (\mathcal{A} - 2 \: \mathcal{B})}{2 \:\sinh \rho} \\
 \partial_{\rho} + \frac{\coth \rho}{2} + \frac{\partial_{\rho} M}{4 \: M} + \frac{l}{\sinh \rho} + \frac{\alpha \: (\mathcal{A} - 2 \: \mathcal{B})}{2 \sinh \rho} & \frac{\iota}{2 \: \sqrt{M}} \Big{(}  1 - M - \alpha \frac{\sin^2 \theta (\rho)}{ \sinh^2 \rho} \Big{)}
\end{pmatrix}\eea

For $\alpha = 1$, the system of equations decouples,

\bea
 \iota \:\mathcal{O}_l = 
\begin{pmatrix} 
 0 & \partial_{\rho} + \frac{\coth \rho}{2} + \frac{\partial_{\rho} M}{4 \: M} - \frac{l}{\sinh \rho} - \frac{(\mathcal{A} - 2 \: \mathcal{B})}{2 \:\sinh \rho} \\
 \partial_{\rho} + \frac{\coth \rho}{2} + \frac{\partial_{\rho} M}{4 \: M} + \frac{l}{\sinh \rho} + \frac{(\mathcal{A} - 2 \: \mathcal{B})}{2 \sinh \rho} & \frac{\iota \: (1 - M )}{\sqrt{M}} \end{pmatrix}.\eea

Take
\bea
\psi_l (\rho) = \begin{bmatrix}
u_l (\rho) \\
v_l (\rho)
\end{bmatrix},
\eea
the equation then  becomes:
\bea
\bigg{(} \partial_{\rho} + \frac{\coth \rho}{2} + \frac{\partial_{\rho} M}{4 \: M} - \frac{l}{\sinh \rho} - \frac{ (\mathcal{A} - 2 \: \mathcal{B})}{2 \:\sinh \rho} \bigg{)} \: v_l (\rho) = 0,  \\
\bigg{(} \partial_{\rho} + \frac{\coth \rho}{2} + \frac{\partial_{\rho} M}{4 \: M} + \frac{l}{\sinh \rho} + \frac{ (\mathcal{A} - 2 \: \mathcal{B})}{2 \:\sinh \rho} \bigg{)} \: u_l (\rho) + \iota \: \frac{1 - M}{\sqrt{M}} \: v_l (\rho) = 0.
\eea
Solving for $v_l (\rho)$ gives,
\bea
v_l (\rho)  = C_1 \: \bigg{(} \sinh \frac{\rho}{2} \bigg{)}^{l-\frac{1}{2}} \:  \bigg{(} \cosh \frac{\rho}{2} \bigg{)}^{-l-\frac{5}{2}}\: \big{(}\cos \theta_0 + \cosh \rho \big{)},
\eea
where $C_1$ is a constant. Using this solution, we can now solve equation for $u_l (\rho)$, 
\bea
u_l^{'} (\rho) + \bigg{(} \frac{\coth \rho}{2} + \frac{\partial_{\rho} M}{4 \: M} + \frac{l}{\sinh \rho} + \frac{ (\mathcal{A} - 2 \: \mathcal{B})}{2 \:\sinh \rho} \bigg{)} \: u_l (\rho) + \iota \: \frac{1 - M}{\sqrt{M}} \: v_l (\rho) = 0.
\eea
The integrating factor for this equation is,
\bea\begin{split}
I (\rho) &= \text{Exp} \bigg{[} \int d \rho \: \bigg{(} \frac{\coth \rho}{2} + \frac{\partial_{\rho} M}{4 \: M} + \frac{l}{\sinh \rho} + \frac{ (\mathcal{A} - 2 \: \mathcal{B})}{2 \:\sinh \rho} \bigg{)} \bigg{]} \\
&= \bigg{(} - \iota \sinh \frac{\rho}{2} \bigg{)}^{l+\frac{1}{2}} \:  \bigg{(} \cosh \frac{\rho}{2} \bigg{)}^{-l-\frac{3}{2}}\:\sqrt{3 + 4 \: \cos \theta_0 \cosh \rho + \cosh (2 \rho)}.
\end{split}\eea
Then, full solution takes the form,
\bea\begin{split}
u_l (\rho) &= \frac{1}{I (\rho)} \bigg{[} \int d \rho\: I (\rho) \: \bigg{(} - \iota \: \frac{1 - M}{\sqrt{M}} \: v_l (\rho)\bigg{)} + C_2 \bigg{]} \\
&=  \bigg{[} C_1 \: \frac{2^{\frac{3}{2} +l} \:\iota \: (2 + 2l + \cosh \rho)\: \big{(} \sinh \frac{\rho}{2} \big{)}^{\frac{1}{2} + 2l} \: \sin^2 \theta_0 }{(3 + 8l + 4 l^2 ) \:\big{(} \cosh \frac{\rho}{2} \big{)}^{\frac{3}{2}} \: \big{(} \sinh \rho \big{)}^l \: \sqrt{3 + 4 \cos \theta_0 \cosh \rho + \cosh (2 \rho)}} \\
&   + C_2 \:\frac{\big{(} \cosh \frac{\rho}{2} \big{)}^{\frac{3}{2} + l} \: \big{(} - \iota \sinh \frac{\rho}{2} \big{)}^{-\frac{1}{2} -l }}{ \sqrt{3 + 4 \cos \theta_0 \cosh \rho + \cosh 2 \rho} } \bigg{]}.
\end{split}\eea

We demand the solution to be regular at the origin, ($\rho = 0$), this fixes $C_2 = 0$ for $l \geq 1/2$ and $C_1 = 0$ for $l \leq - 1/2$. \\

For $l \leq -1/2$,
\bea
u_l^{-} (\rho) = C_2 \:\frac{\big{(} \cosh \frac{\rho}{2} \big{)}^{\frac{3}{2} + l} \: \big{(} - \iota \sinh \frac{\rho}{2} \big{)}^{-\frac{1}{2} -l }}{ \sqrt{3 + 4 \cos \theta_0 \cosh \rho + \cosh 2 \rho} },
\eea
and for $l \geq 1/2$,
\bea
u_l^{+} (\rho) = C_1 \: \frac{2^{\frac{3}{2} +l} \:\iota \: (2 + 2l + \cosh \rho)\: \big{(} \sinh \frac{\rho}{2} \big{)}^{\frac{1}{2} + 2l} \: \sin^2 \theta_0 }{(3 + 8l + 4 l^2 ) \:\big{(} \cosh \frac{\rho}{2} \big{)}^{\frac{3}{2}} \: \big{(} \sinh \rho \big{)}^l \: \sqrt{3 + 4 \cos \theta_0 \cosh \rho + \cosh (2 \rho)}},
\eea
\bea
v_l^{+} (\rho) = C_1 \: \bigg{(} \sinh \frac{\rho}{2} \bigg{)}^{l-\frac{1}{2}} \:  \bigg{(} \cosh \frac{\rho}{2} \bigg{)}^{-l-\frac{5}{2}}\: \big{(}\cos \theta_0 + \cosh \rho \big{)}.
\eea

Let us now present the relevant formulas in this case $(m = 0)$:
\bea
 \big{(} (m + V)^2 - m^2 - W^2 \big{)} = 0 \eea
\bea
\hat{\zeta}_{\mathcal{O}} (0) &=& - \frac{1}{2} \int_0^{\infty} d\rho \: \sinh \rho \: \big{(} (m + V)^2 - m^2 - W^2 \big{)} = 0
\eea
and
\bea
 \int_0^{\infty} d\rho \: \sinh \rho \: \ln \bigg{(} \frac{\sinh \rho}{2}\bigg{)}\: \big{(} (m + V)^2 - m^2 - W^2 \big{)} = 0
\eea
together with
\bea
\int_0^{\infty} d\rho \: \sinh \rho \: W^2 =  - \frac{1}{8}  \theta_0 \: \sin \theta_0 + \frac{1}{2} \: \sin^2 \frac{\theta_0}{2} 
\eea

The Weyl anomaly contribution in this case,
\bea\begin{split}
&\frac{1}{4\pi} \: \int d^2 \sigma \sqrt{g} \ln M \: \bigg{[} (m + V)^2 - W^2 + \frac{1}{12} R - \frac{1}{24} \nabla^2 \ln M \bigg{]} \\
&= \frac{1}{4\pi} \: \int d^2 \sigma \sqrt{g} \ln M \: \bigg{[} \frac{1}{12} R - \frac{1}{24} \nabla^2 \ln M \bigg{]}\\
&=- \frac{1}{4} \sin^2 \frac{\theta_0}{2} - \frac{1}{6} \log \cos \frac{\theta_0}{2}
\end{split} \eea

Total contribution from these last two cases:
\bea\begin{split}
&\ln \bigg{(} \frac{\text{det} \:O}{\text{det} \:O^{\text{free}}}\bigg{)} = - 2 \: \bigg{(}\frac{1}{2} \bigg{)}^2 \int_0^{\infty} d \rho \: \frac{(\mathcal{A} + 2 \mathcal{B})^2}{\sinh \rho} - 2 \: \bigg{(} - \frac{1}{2} \bigg{)}^2 \int_0^{\infty} d \rho \: \frac{(\mathcal{A} + 2 \mathcal{B})^2}{\sinh \rho} \\
& - 2 \: \bigg{(}\frac{1}{2} \bigg{)}^2 \int_0^{\infty} d \rho \: \frac{(\mathcal{A} - 2 \mathcal{B})^2}{\sinh \rho} - 2 \: \bigg{(} - \frac{1}{2} \bigg{)}^2 \int_0^{\infty} d \rho \: \frac{(\mathcal{A} - 2 \mathcal{B})^2}{\sinh \rho} - 4 \: \int_0^{\infty} d \rho \sinh \rho \: W^2 \\
&= - \int_0^{\infty} d \rho \: \frac{\mathcal{A}^2}{\sinh \rho} - 4 \: \int_0^{\infty} d \rho \: \frac{\mathcal{B}^2}{\sinh \rho} - 4 \: \int_0^{\infty} d \rho \sinh \rho \: W^2
\end{split}\eea
Thus,
\bea\begin{split}
&\ln \bigg{(} \frac{\text{det} \:O}{\text{det} \:O^{\text{free}}}\bigg{)} = \frac{1}{2} \theta_0 \: \sin \theta_0 + \sin^2 \frac{\theta_0}{2} + 6 \: \log \bigg{(} \cos \frac{\theta_0}{2}\bigg{)}
\end{split}\eea

\subsection{One-loop effective action}

The total zeta-function at the origin
\bea
\hat{\zeta}_{\text{tot}} (0) = 2 \: \hat{\zeta}_{\mathcal{O}_1} (0) + \hat{\zeta}_{\mathcal{O}_{2+}} (0) + \hat{\zeta}_{\mathcal{O}_{2-}} (0) + 2 \: \hat{\zeta}_{\mathcal{O}_{3+}} (0) + 2\: \hat{\zeta}_{\mathcal{O}_{3-}} (0) - 2 \hat{\zeta}_{\mathcal{O}_{+} } (0) - 2 \hat{\zeta}_{\mathcal{O}_{-} } (0)
\eea
where $\mathcal{O}_{\pm}$ are $N =4$ fermionic operators.
\bea
4 \: \big{(} (1 + V)^2 - W^2 - 1\big{)} - 2 \: V_2 - 4 \: V_3 = \nabla^2 \: \ln M
\eea
which vanishes when integrated,
\bea
\int_0^{\infty} d \rho \sinh \rho\: \nabla^2 \: \ln M = \sinh \rho \: \partial_{\rho} \ln M \vert_{0}^{\infty} = 0, \:\:\: \:\:\:\: \hat{\zeta}_{\text{tot}} (0)  = 0.
\eea

The contributions from gauge field, seen to vanish:
\bea\begin{split}
 &1 \times (1)^2 \mathcal{A}^2 + 1  \times(-1)^2 \mathcal{A}^2 + 2 \times (1)^2 \mathcal{B}^2 + 2  \times(-1)^2 \mathcal{B}^2 - 2  \times(\frac{1}{2})^2 \mathcal{A}^2 - 2  \times(- \frac{1}{2})^2 \mathcal{A}^2 \\
&- 1 \times (\frac{1}{2})^2 \times (\mathcal{A} + 2  \mathcal{B})^2 - 1  \times(- \frac{1}{2})^2 (\mathcal{A} + 2  \mathcal{B})^2  - 1  \times(\frac{1}{2})^2 (\mathcal{A} - 2 \mathcal{B})^2 - 1  \times(- \frac{1}{2})^2 (\mathcal{A} - 2 \mathcal{B})^2 \\
&= 0
\end{split}\eea

The contribution from $W^2$ term in the fermionic potenital,
\bea\begin{split}
W^2: \:\:\:\:& - 4 \times \int_0^{\infty} d\rho \sinh \rho \: \bigg{(} \frac{\sin^2 \theta (\rho)}{\sqrt{M} \: \sinh^2 \rho}\bigg{)}^2 - 4 \times \int_0^{\infty} d\rho \sinh \rho \: \bigg{(} - \frac{\sin^2 \theta (\rho)}{2 \: \sqrt{M} \: \sinh^2 \rho}\bigg{)}^2 \\
&= \frac{5}{2} \theta_0 \: \sin \theta_0 - 10 \sin^2 \frac{\theta_0}{2}
\end{split}\eea

Weyl anomaly contribution:
\begin{itemize}
\item Potential and mass terms for the $\frac{1}{6}-$BPS operators
\bea
4\: \big{(} (1 + V)^2 - W^2 \big{)} + 2 \times 1 - 2 \times 2 - 2 \times V_2 - 4 \times V_3 = - R + \nabla^2 \ln M
\eea
\item Curvature and conformal terms
\bea
\bigg{(} 8 \times \bigg{(} \frac{1}{12} \bigg{)} - 8 \times \bigg{(} - \frac{1}{6} \bigg{)} R + \nabla^2 \ln M \bigg{(} 8 \times \bigg{(} - \frac{1}{24} \bigg{)} - 8 \times \bigg{(} \frac{1}{12} \bigg{)} \bigg{)} = 2 R - \nabla^2 \ln M.
\eea

\end{itemize}

Total contribution from Weyl anomaly,
\bea
\text{anomaly}: \;\;\;\;\;\;\; \frac{1}{4\pi} \int d^2 \sigma \sqrt{g} R \ln M = - \bigg{(} \theta_0 \sin \theta_0 + 4 \: \cos^2 \frac{\theta_0}{2} \: \ln \cos \frac{\theta_0}{2} \bigg{)}
\eea

The contribution from $\ln \: (\sinh \rho)$ integrals involve the same combination of potentials as $\hat{\zeta}_{\text{tot}} (0)$, which when added to the Weyl anomaly gives,
\bea
\text{anomaly} + \ln \sinh \rho : \:\:\:\:\: \int_0^{\infty} \: d \rho \: \sinh \rho \: \bigg{(} \frac{1}{2} R\: \ln M + \ln \bigg{(} \frac{\sinh \rho}{2}\bigg{)}  \nabla^2 \: \ln M \bigg{)} = - 2 \ln \cos \frac{ \theta_0}{2}.
\eea

\bea\begin{split}
 \Delta \Gamma^{\text{1-loop}}_{\text{effective}} (\theta_0) &=  \frac{5}{4} {\theta_0}  \sin \theta_0 - 5 \: \sin^2 \frac{\theta_0}{2} + 2 \: \ln \cos \frac{\theta_0}{2}   \\
& + 2 \: \ln \: \Gamma \bigg{(} \cos^2 \frac{\theta_0}{2} \bigg{)}   - \ln \big{(} \Gamma (\cos \theta_0)\big{)}  -  \ln \big{(} \cos \theta_0\big{)} \\
&=  \frac{1}{4} \theta_0^2 + O (\theta_0^4)
\end{split}\eea

This result does not agree with the field theory expectation. Although our regularization is diffeomorphic invariant there might be ambiguities that need to be understood better. At the moment we can track the discrepancy between the two methods to an ambiguity in the treatment of the $m=0$ fermionic modes, we will return to this question elsewhere. It seems that a more expeditious way to get at the exact answer might follow the approach of  \cite{Cagnazzo:2017sny,Medina-Rincon:2018wjs} who mapped the spectral problems from the disk to the cylinder with the incorporation of an explicit diffeomorphic invariant cutoff; we hope to report on such explorations in an upcoming publication. 


\section{Conclusions}\label{Sec:Conclusions}

In this manuscript we have discussed in detail the construction of the quadratic fluctuations for the string configuration dual to the general latitude Wilson loop in ABJM theory. We have paid particular attention to the various symmetries of the configurations and shown how they serve as a guiding avatar in the structure of fluctuations. At the semiclassical level the computation of the one-loop effective action is equivalent to the computation of determinants. We employed  two methods for computing such determinants. The perturbative heat kernel method has lead to agreement with the expected field theory answer in the limit of small latitude angle. The $\zeta$-function regularization method is non-perturbative but does not seem to lead to the expected field theory answer as it stands.  We have previously developed the $\zeta$-function approach in  \cite{Aguilera-Damia:2018rjb} and applied it to the ${\cal N}=4$ context in  \cite{Aguilera-Damia:2018twq} motivated by the goal of constructing a regularization that is explicitly diffeomorphic invariant. The key new ingredient in this work that introduces extra ambiguities with respect to our earlier efforts is the fact that some of the modes correspond to massless fermions. The situation is not completely satisfactory but sheds light on deficiencies and advantages of various methods used to tackle questions of precision holography with Wilson loops. For example, some of the puzzles we face were confronted in the realm of ${\cal N}=4$ SYM and paved the way for a hybrid approach leading to perfect matching with the field theory answer in   \cite{Cagnazzo:2017sny} where the computations of the determinants was mapped from the disk to the cylinder. We hope to revisit our computations using a similar approach. 

One interesting property of the duality pair we discuss is that it admits two very natural limits. Here we focused on the `t Hooft limit where $\lambda=N/k$ is kept fixed as $N$ is taken very large. It would be interesting to explore the M-theory limit, where $k$ is kept fixed,  beyond the leading order as well; some preliminary results were reported in \cite{Sakaguchi:2010dg}. Exploring quantum corrections in this context might ultimately shed light on various intricate quantum properties of M2 branes. 

 It would also  be interesting to explore similar issues for Wilson loops in higher dimensional representations. Classical results were presented in \cite{Drukker:2008zx,Rey:2008bh}; at the quantum level some preliminary results have been presented in \cite{Muck:2016hda} for the gravity configurations and a sub-leading analysis of the matrix model was presented in \cite{Cookmeyer:2016dln}. The prospects for precision holography in this case are improved due to the fact that the corresponding quadratic fluctuations live in the odd-dimensional world-volumes of the corresponding D2 and D6 branes \cite{Muck:2016hda}. Heat kernel techniques are considerably simplified in odd-dimensional spaces since the contributions arise exclusively from zero or boundary modes. 

Recently, in the case of ${\cal N}=4$ SYM, the expectation value of the  $\frac{1}{2}$-BPS Wilson loop has been computed on the gravity side by taking the ratio of two of the limits of the latitude string \cite{Medina-Rincon:2018wjs}. We hope that a similar analysis in the case of ABJM Wilson loops will shed light on various aspects of precision holography in IIA, our work provides most of the required ingredients.

\section*{Acknowledgments}
We thank D. Trancanelli and E. Vescovi. LPZ, VR and GAS thank ICTP for providing  hospitality at various stages. 
AF was supported by Fondecyt \# 1160282.  LPZ and VR are partially supported by the US Department of Energy under Grant No. DE-SC0017808 –{\it  Topics in the AdS/CFT Correspondence: Precision tests with Wilson loops, quantum black holes and dualities} and Grant No. DE-SC0007859.  GAS and JAD are supported by CONICET and grants PICT
2012-0417, PIP0595/13, X648 UNLP, PIP 0681 and PI {\it B\'usqueda de nueva F\'isica}. 
\appendix
\section{Conventions}\label{App:conventions}
Ten-dimensional target-space indices are denoted by $m,n,\ldots$, two-dimensional world-sheet indices are $a,b,\ldots$, while the directions orthogonal to the string are represented by $i,j,\ldots$. All corresponding tangent space indices are underlined. 

In Euclidean signature the Dirac matrices satisfy
\begin{empheq}{alignat=7}
	\Gamma_{\underline{m}}^{\dagger}&=\Gamma_{\underline{m}}\,,
	&\qquad
	\Gamma_{\underline{m}}^2&=1\,,
\end{empheq}
and the chirality matrix is
\begin{empheq}{alignat=7}
	\Gamma_{11}&\equiv-i\Gamma_{\underline{0123456789}}\,,
	&\qquad
	\Gamma_{11}^{\dagger}&=\Gamma_{11}\,,
	&\qquad
	\Gamma_{11}^2&=1\,.
\end{empheq}
The charge conjugation intertwiners $C_{\pm}$ are such that
\begin{empheq}{alignat=7}
	C_{\pm}\Gamma_{\underline{m}}C_{\pm}^{-1}&=\pm\Gamma_{\underline{m}}^T\,,
	&\qquad
	C_{\pm}\Gamma_{11}C_{\pm}^{-1}&=-\Gamma_{11}^T\,,
	&\qquad
	C_{\pm}^T&=\pm C_{\pm}\,.
\end{empheq}
Majorana spinors are defined as
\begin{empheq}{alignat=7}
	\psi^TC_{\pm}&=\psi^{\dagger}
	&\qquad\Leftrightarrow\qquad	
	\psi^*&=\pm C_{\pm}\psi\,.
\end{empheq}

In Lorentzian signature we have
\begin{empheq}{alignat=7}
	\Gamma_{\underline{m}}^{\dagger}&=\Gamma_{\underline{0}}\Gamma_{\underline{m}}\Gamma_{\underline{0}}\,,
	&\qquad
	\Gamma_{\underline{0}}^2&=-1\,,
	&\qquad
	\Gamma_{\underline{m}\neq0}^2&=1\,.
\end{empheq}
and the chirality matrix reads
\begin{empheq}{alignat=7}
	\Gamma_{11}&\equiv\Gamma_{\underline{0123456789}}\,,
	&\qquad
	\Gamma_{11}^{\dagger}&=\Gamma_{11}\,,
	&\qquad
	\Gamma_{11}^2&=1\,.
\end{empheq}

\section{Geometric Data}\label{app: Geometric Data}

In this appendix we collect all the geometric formulae necessary to compute the spectrum of excitations of the $1/6$-BPS string.

We start by writing the target space fields. The Euclidean $AdS_4$ ($EAdS_4$) metric is written as an $\mathds{H}_2\times S^1$ foliation,
\begin{empheq}{align}\label{AdS5 metric}
	ds^2_{EAdS_4}&=\cosh^2u\left(\sinh^2\rho\,d\psi^2+d\rho^2\right)+\sinh^2u\,d\phi^2+du^2\,,
\end{empheq}
with $u\geq0$, $\rho\geq0$, $\psi\sim\psi+2\pi$ and $\phi\sim\phi+2\pi$. The metric on $\mathds{CP}^3$ is taken to be
\begin{empheq}{align}
	ds^2_{\mathds{CP}^3}&=\frac{1}{4}\left[d\alpha^2+\cos^2\frac{\alpha}{2}\left(d\vartheta_1^2+\sin^2\vartheta_1\,d\varphi_1^2\right)+\sin^2\frac{\alpha}{2}\left(d\vartheta_2^2+\sin^2\vartheta_2\,d\varphi_2^2\right)\right.
	\nonumber\\
	&\phantom{=}\left.+\cos^2\frac{\alpha}{2}\sin^2\frac{\alpha}{2}\left(d\chi-\left(1-\cos\vartheta_1\right)\,d\varphi_1+\left(1-\cos\vartheta_2\right)\,d\varphi_2\right)^2\right]\,,
\end{empheq}
where $0\leq\alpha\leq\pi$, $0\leq\vartheta_1\leq\pi$, $0\leq\vartheta_1\leq\pi$, $\varphi_1\sim\varphi_1+2\pi$, $\varphi_2\sim\varphi_2+2\pi$ and $\chi\sim\chi+4\pi$. The full $EAdS_4\times\mathds{CP}^3$ metric with radius $L$ is then
\begin{empheq}{alignat=7}
	ds^2&=L^2\left(ds^2_{EAdS_4}+4\,ds^2_{\mathds{CP}^3}\right)\,.
\end{empheq}
The other background fields read
\begin{empheq}{alignat=7}
	e^{\Phi}&=\frac{2L}{k}\,,
	&\qquad
	F_{(4)}&=-\frac{3ikL^2}{2}\textrm{vol}\left(AdS_4\right)\,,
	&\qquad
	F_{(2)}&=\frac{k}{4}J\,,
\end{empheq}
where
\begin{empheq}{alignat=7}
	\textrm{vol}\left(AdS_4\right)&=\cosh^2u\sinh u\sinh\rho\,d\psi\wedge d\rho\wedge du\wedge d\phi\,,
	\\
	J&=-2\cos\frac{\alpha}{2}\sin\frac{\alpha}{2}\,d\alpha\wedge\left(d\chi-\left(1-\cos\vartheta_1\right)\,d\varphi_1+\left(1-\cos\vartheta_2\right)\,d\varphi_2\right)
	\\\nonumber
	&-2\cos^2\frac{\alpha}{2}\sin\vartheta_1\,d\vartheta_1\wedge d\varphi_1-2\sin^2\frac{\alpha}{2}\sin\vartheta_2\,d\vartheta_1\wedge d\varphi_2\,.
\end{empheq}
The factor of $i$ in $F_{(4)}$ is due to the Euclidean continuation. The 2-form is proportional to the Kahler form in $\mathbb{CP}^3$.

Target space indices are labeled by $m,n,......,$ worldvolume indices are $a,b,....,$  directions orthogonal to the string are denoted by $i,j,.....$ The corresponding target space indices are underlined.

The choice of adapted $EAdS_4 \times \mathbb{CP}^3$ vielbein $E^{\underline{m}} = \big{(} E^{\underline{a}}, E^{\underline{i}}\big{)}$ is
\begin{empheq}{alignat=7}\label{10D Vielbein}
	E^{\underline{0}}&=L\,A^{-\frac{1}{2}}\left(\cosh^2u\sinh^2\rho\,\dot{\psi}\,d\psi+\cos^2\frac{\alpha}{2}\sin^2\vartheta_1\,\dot{\varphi}_1\,d\varphi_1\right)\,,
	\nonumber\\
	E^{\underline{1}}&=L\,B^{-\frac{1}{2}}\left(\cosh^2u\,\rho'\,d\rho+\cos^2\frac{\alpha}{2}\,\vartheta_1'\,d\vartheta_1\right)\,,
	\nonumber\\
	E^{\underline{2}}&=L\,du\,,
	\nonumber\\
	E^{\underline{3}}&=L\,\sinh u\,d\phi\,,
	\nonumber\\
	\left(
	\begin{array}{c}
		E^{\underline{4}}
		\\
		E^{\underline{5}}
	\end{array}
	\right)
	&=
	\left(
	\begin{array}{rr}
		\cos\Delta & \sin\Delta
		\\
		-\sin\Delta & \cos\Delta
	\end{array}
	\right)
	\left(
	\begin{array}{c}
		L\,B^{-\frac{1}{2}}\cosh u\cos\frac{\alpha}{2}\left(\rho'\,d\vartheta_1-\vartheta_1'\,d\rho\right)
		\\
		L\,A^{-\frac{1}{2}}\cosh u\sinh\rho\cos\frac{\alpha}{2}\sin\vartheta_1\left(\dot{\psi}\,d\varphi_1-\dot{\varphi}_1\,d\psi\right)
	\end{array}
	\right)\,,
	\nonumber\\
E^{\underline{6}}&=L\,\sin\frac{\alpha}{2}\,d\vartheta_2\,,
	\\\nonumber
	E^{\underline{7}}&=L\,\sin\frac{\alpha}{2}\sin\vartheta_2\,d\varphi_2\,,
	\\\nonumber
	E^{\underline{8}}&=L\,d\alpha\,,
	\\\nonumber
	E^{\underline{9}}&=L\,\cos\frac{\alpha}{2}\sin\frac{\alpha}{2}\left(d\chi-\left(1-\cos\vartheta_1\right)\,d\varphi_1+\left(1-\cos\vartheta_2\right)\,d\varphi_2\right)
\end{empheq}
where
\begin{empheq}{alignat=7}
	A(u,\rho,\alpha,\vartheta_1)&=\cosh^2u\sinh^2\rho\,\dot{\psi}^2+\cos^2\frac{\alpha}{2}\sin^2\vartheta_1\,\dot{\varphi_1}^2\,,
	\nonumber\\
	B(u,\rho,\alpha,\vartheta_1)&=\cosh^2u\,\rho'^2+\cos^2\frac{\alpha}{2}\,\vartheta_1'^2\,.
\end{empheq}
Here $\dot{\psi}=\frac{d\psi}{d\tau}$ and $\dot{\varphi}_1=\frac{d\varphi_1}{d\tau}$ are constant numbers while $\rho'=\frac{d\rho}{d\sigma}$ and $\vartheta_1'=\frac{d\vartheta_1}{d\sigma}$ are understood as functions of $\rho$ and $\vartheta_1$, respectively. Also, $\Delta$ is an arbitrary function of $\psi$ and $\varphi_1$ describing and $SO(2)$ rotation of the canonical frames and it is to be chosen at our convenience. The standard $EAdS_4\times\mathds{CP}^3$ vielbein is recovered for $\rho'=1$, $\vartheta_1'=0$, $\dot{\psi}=1$ and $\dot{\varphi_1}=0$, and $\Delta=0$. For the $1/6$-BPS solution, $\rho'=-\sinh\rho$, $\vartheta_1'=-\sin\vartheta_1$ and $\dot{\psi}=\dot{\varphi}_1=1$. The standard and the adapted vielbein are related by the local Lorentz transformation
\begin{empheq}{alignat*=7}
	S&=e^{\Delta J_{\underline{45}}}e^{aJ_{\underline{05}}}e^{bJ_{\underline{14}}}\,,
\end{empheq}
where
\begin{empheq}{alignat=7}
	\cos a&=\frac{\cosh u\sinh\rho\,\dot{\psi}}{\sqrt{A}}\,,
	&\qquad
	\sin a&=\frac{\cos\frac{\alpha}{2}\sin\vartheta_1\,\dot{\varphi}_1}{\sqrt{A}}\,,
	\\
	\cos b&=\frac{\cosh u\,\rho'}{\sqrt{B}}\,,
	&\qquad
	\sin b&=\frac{\cos\frac{\alpha}{2}\,\vartheta_1'}{\sqrt{B}}\,.
\end{empheq}
Notice that for $\rho'=-\dot{\psi}\sinh\rho$ and $\vartheta_1'=-\dot{\varphi}_1\sin\vartheta_1$ we have
\begin{empheq}{alignat=7}
	b&=a+\pi\,.
\end{empheq}
For reasons to be explained below, we shall set $\Delta$ such that $\Delta=\tau$ on the worldsheet. 

The adapted vielbein has the desired property that upon taking the pullback onto the worldsheet
\begin{empheq}{alignat=7}
	P[E^{\underline{a}}]&=e^{\underline{a}}\,,
	&\qquad
	\underline{a}&=0,1\,,
	\\
	P[E^{\underline{i}}]&=0\,,
	&\qquad
	\underline{i}&=2,\ldots,9\,,
\end{empheq}
where 
\begin{empheq}{alignat=7}
	e^{\underline{0}}&=\sqrt{A}\,d\tau\,,
	&\quad
	e^{\underline{1}}&=\sqrt{A}\,d\sigma\,,
\end{empheq}
is a vielbein for the induced geometry
\begin{empheq}{alignat=7}
	ds^2_{\textrm{ind}}&=A\left(d\tau^2+d\sigma^2\right)\,.
\end{empheq}
The conformal factor reads
\begin{empheq}{alignat=7}
	A(\sigma)&=\sinh^2\rho+\sin^2\vartheta_1&&=\frac{1}{\sinh^2\sigma}+\frac{1}{\cosh^2\left(\sigma+\sigma_0\right)}\,.
\end{empheq}

The worldsheet spin connection, the extrinsic curvature and the normal bundle gauge fields are given by, respectively,
\begin{empheq}{alignat=5}
	w^{\underline{ab}}&=P[\Omega^{\underline{ab}}]\,,
	&\qquad
	 H^{\underline{i}}_{\phantom{\underline{i}}ab}&=P[\Omega^{\underline{i}}_{\phantom{\underline{i}}\underline{a}}]_ae^{\underline{a}}_{\phantom{\underline{a}}b}\,,
	&\qquad
	\mathcal{A}^{\underline{ij}}&=P[\Omega^{\underline{ij}}]\,,
\end{empheq}
where $\Omega^{\underline{mn}}$ is the target space spin connection. For the $\frac{1}{6}$-BPS string we find
\begin{empheq}{alignat=7}
	w^{\underline{01}}&=\frac{A'}{2A}\,d\tau\equiv w\,d\tau\,,
	\\
	\mathcal{A}^{\underline{45}}&=\frac{\cosh\rho\cos\vartheta_1+1}{\cosh\rho+\cos\vartheta_1}\,d\tau-P[d\Delta]=\left(\tanh(2\sigma+\sigma_0)-\dot{\Delta}\right)\,d\tau\,,
	\\
	\mathcal{A}^{\underline{67}}&=\frac{1}{2}\left(1-\cos\vartheta_1\right)\,d\tau=\frac{1}{2}\left(1-\tanh(\sigma+\sigma_0)\right)\,d\tau\,,
	\\
	\mathcal{A}^{\underline{89}}&=\frac{1}{2}\left(1-\cos\vartheta_1\right)\,d\tau=\frac{1}{2}\left(1-\tanh(\sigma+\sigma_0)\right)\,d\tau\,,
\end{empheq}
and
\begin{empheq}{alignat=7}
	H^{\underline{4}\phantom{a}b}_{\phantom{\underline{5}}a}&=\frac{m}{\sqrt{A}}\left(
	\begin{array}{rr}
		-\cos\Delta & \sin\Delta
		\\
		\sin\Delta & \cos\Delta
	\end{array}
	\right)\,,
	&\qquad
	H^{\underline{5}\phantom{a}b}_{\phantom{\underline{6}}a}&=\frac{m}{\sqrt{A}}\left(
	\begin{array}{rr}
		\sin\Delta & \cos\Delta
		\\
		\cos\Delta & -\sin\Delta
	\end{array}
	\right)\,,
\end{empheq}
where
\begin{empheq}{alignat=7}
	m&=\frac{\sinh\rho\sin\vartheta_1}{\cosh\rho-\cos\vartheta_1}=\frac{1}{\cosh\left(2\sigma+\sigma_0\right)}\,.
\end{empheq}

For the purpose of computing the spectrum of fluctuations we will chose $\Delta$ such that
\begin{empheq}{alignat=7}
	P[d\Delta]&=d\tau
	&\qquad
	\left(\textrm{e.g. }\Delta=\psi\right)\,.
\end{empheq}
The reason for this choice is that the gauge fields
\begin{empheq}{alignat=7}
	\mathcal{A}&\equiv\mathcal{A}^{\underline{45}}=\left(\tanh(2\sigma+\sigma_0)-1\right)\,d\tau\,,
	\\
	\mathcal{B}&\equiv\mathcal{A}^{\underline{67}}=\mathcal{A}^{\underline{89}}=\frac{1}{2}\left(1-\tanh(\sigma+\sigma_0)\right)\,d\tau\,,
\end{empheq}
are then regular at the center of the disk $\sigma\rightarrow\infty$, where the 1-form $d\tau$ is not well defined. Indeed\footnote{Near the center of the disk the metric becomes $ds^2=dr^2+r^2d\tau^2$, with $r=2e^{-\sigma}\sqrt{1+e^{-2\sigma_0}}$. Regularity of the gauge fields requires that $d\tau$ be multiplied by $r^n$, $n\geq2$, as $r\rightarrow0$.} $\mathcal{A}\sim e^{-4\sigma}$ and $\mathcal{B}\sim e^{-2\sigma}$ as $\sigma\rightarrow\infty$ . They also vanish in the $1/2$-BPS limit $\sigma_0\rightarrow\infty$. Notice that
\begin{empheq}{alignat=7}
	w-\mathcal{A}&=1-\cosh\rho-\cos\vartheta_1\,,
	&\qquad
	\partial_{\sigma}\mathcal{A}&=2m^2\,,
	&\qquad
	\partial_{\sigma}\mathcal{B}&=-\frac{1}{2}\sin^2\vartheta_1\,.
\end{empheq}
These relations  prove to be useful when casting the equations of motion in a simple form. 

Finally, the contractions involving the Riemann tensor that we need are
\begin{empheq}{alignat=7}
	\delta^{\underline{ab}}R_{\underline{aibj}}&=\left\{
	\begin{array}{cl}
		{\displaystyle-\frac{2\sinh^2\rho}{A}} & \underline{i}=\underline{j}=\underline{2},\underline{3}
		\\\\
		{\displaystyle\frac{\sin^2\vartheta_1}{2A}} & \underline{i}=\underline{j}=\underline{6},\underline{7},\underline{8},\underline{9}
		\\\\
		{\displaystyle0} & \textrm{otherwise}
	\end{array}
	\right.\,.
\end{empheq}

It is useful to invert the vielbein in order to write the RR fields that enter in the spinor action and Killing equation. We will set $\Delta=0$ in this computation and then argue that some of the results do not depend on $\Delta$. For generality we leave $\rho'$, $\vartheta_1'$, $\dot{\psi}$ and $\dot{\varphi}_1$ arbitrary. We have,
\begin{empheq}{alignat=7}
	\cosh u\sinh\rho\,d\psi&=\frac{1}{L\sqrt{A}}\left(\cosh u\sinh\rho\,\dot{\psi}\,E^{\underline{0}}-\cos\frac{\alpha}{2}\sin\vartheta_1\,\dot{\varphi}_1\,E^{\underline{5}}\right)\,,
	\\
	\cos\frac{\alpha}{2}\sin\vartheta_1\,d\varphi_1&=\frac{1}{L\sqrt{A}}\left(\cos\frac{\alpha}{2}\sin\vartheta_1\,\dot{\varphi}_1\,E^{\underline{0}}+\cosh u\sinh\rho\,\dot{\psi}\,E^{\underline{5}}\right)\,,
	\\
	\cosh u\,d\rho&=\frac{1}{L\sqrt{B}}\left(\cosh u\,\rho'\,E^{\underline{1}}-\cos\frac{\alpha}{2}\,\vartheta_1'\,E^{\underline{4}}\right)\,,
	\\
	\cos\frac{\alpha}{2}\,d\vartheta_1&=\frac{1}{L\sqrt{B}}\left(\cos\frac{\alpha}{2}\,\vartheta_1'\,E^{\underline{1}}+\cosh u\,\rho'\,E^{\underline{4}}\right)\,.
\end{empheq}
These relations imply that
\begin{empheq}{alignat=7}
	F_{(4)}&=-\frac{3ik}{2L^2\sqrt{AB}}\left(\cosh u\sinh\rho\,\dot{\psi}\,E^{\underline{0}}-\cos\frac{\alpha}{2}\sin\vartheta_1\,\dot{\varphi}_1\,E^{\underline{5}}\right)\wedge\left(\cosh u\,\rho'\,E^{\underline{1}}-\cos\frac{\alpha}{2}\,\vartheta_1'\,E^{\underline{4}}\right)
	\\
	&\phantom{=}\wedge E^{\underline{2}}\wedge E^{\underline{3}}\,,
	\\
	F_{(2)}&=-\frac{k}{2L^2\sqrt{AB}}\left(-\left(\cos\frac{\alpha}{2}\sin\vartheta_1\,\dot{\varphi}_1\,E^{\underline{0}}+\cosh u\sinh\rho\,\dot{\psi}\,E^{\underline{5}}\right)\wedge\left(\cos\frac{\alpha}{2}\,\vartheta_1'\,E^{\underline{1}}+\cosh u\,\rho'\,E^{\underline{4}}\right)\right.
	\\
	&\left.+\sqrt{AB}\left(E^{\underline{6}}\wedge E^{\underline{7}}+E^{\underline{8}}\wedge E^{\underline{9}}\right)\right)\,,
\end{empheq}
which allows us to compute the following quantities needed for the fermionic fluctuations:
\begin{empheq}{alignat=7}
	\slashed{F}_{(4)}&=-\frac{3ik}{2L^2\sqrt{AB}}\left(\cosh u\sinh\rho\,\dot{\psi}\,\Gamma^{\underline{0}}-\cos\frac{\alpha}{2}\sin\vartheta_1\,\dot{\varphi}_1\,\Gamma^{\underline{5}}\right)\left(\cosh u\,\rho'\,\Gamma^{\underline{1}}-\cos\frac{\alpha}{2}\,\vartheta_1'\,\Gamma^{\underline{4}}\right)\Gamma^{\underline{23}}
	\\
	\slashed{F}_{(2)}&=-\frac{k}{2L^2\sqrt{AB}}\left(-\left(\cos\frac{\alpha}{2}\sin\vartheta_1\,\dot{\varphi}_1\,\Gamma^{\underline{0}}+\cosh u\sinh\rho\,\dot{\psi}\,\Gamma^{\underline{5}}\right)\left(\cos\frac{\alpha}{2}\,\vartheta_1'\,\Gamma^{\underline{1}}+\cosh u\,\rho'\,\Gamma^{\underline{4}}\right)\right.
	\\
	&\left.+\sqrt{AB}\left(\Gamma^{\underline{67}}+\Gamma^{\underline{89}}\right)\right)\,,
\end{empheq}
and
\begin{empheq}{alignat=7}
	\frac{1}{8}e^{\Phi}\Gamma^{\underline{a}}\slashed{F}_{(4)}\Gamma_{\underline{a}}&=\frac{3i}{4L\sqrt{AB}}\left(\cosh^2u\sinh\rho\,\rho'\,\dot{\psi}\,\Gamma^{\underline{01}}+\cos^2\frac{\alpha}{2}\sin\vartheta_1\,\vartheta_1'\,\dot{\varphi}_1\Gamma^{\underline{45}}\right)\Gamma^{\underline{23}}\,,
	\\
	\frac{1}{8}e^{\Phi}\Gamma^{\underline{a}}\slashed{F}_{(2)}\Gamma_{11}\Gamma_{\underline{a}}&=\frac{1}{4L\sqrt{AB}}\left(\cos^2\frac{\alpha}{2}\sin\vartheta_1\vartheta_1'\,\dot{\varphi}_1\,\Gamma^{\underline{01}}+\cosh^2u\sinh\rho\,\rho'\,\dot{\psi}\,\Gamma^{\underline{45}}\right.
	\\
	&\left.+\sqrt{AB}\left(\Gamma^{\underline{67}}+\Gamma^{\underline{89}}\right)\right)\Gamma_{11}\,.
\end{empheq}
On the $\frac{1}{6}$-BPS solution the fermionic mass term becomes
\begin{empheq}{alignat*=7}
	\frac{1}{8}e^{\Phi}\Gamma^{\underline{a}}\left(\slashed{F}_{(2)}\Gamma_{11}+\slashed{F}_{(4)}\right)\Gamma_{\underline{a}}&=\frac{1}{4LA}\left(\sinh^2\rho\left(-3i\Gamma^{\underline{0123}}+\left(-\Gamma^{\underline{45}}+\Gamma^{\underline{67}}+\Gamma^{\underline{89}}\right)\Gamma_{11}\right)\right.
	\\
	&\left.+\sin^2\vartheta_1\left(-3i\Gamma^{\underline{2345}}+\left(-\Gamma^{\underline{01}}+\Gamma^{\underline{67}}+\Gamma^{\underline{89}}\right)\Gamma_{11}\right)\right)
\end{empheq}
Notice that only quantities that are invariant under rotations in the $4-5$, $6-7$ and $8-9$ planes appear in the last two expressions. Therefore, these are also valid for arbitrary choices of $\Delta$. In particular, they hold in the rotated frame where the connections are regular.

\section{Regular gauge fields and spinors}\label{app:regularity} 
The discussion about the regularity of the gauge fields is important because it is coupled to the periodicity of the fields. On general grounds, we expect regular bosonic/fermionic fields to be periodic/anti-periodic. Since a gauge transformation can change the periodicity of the fields, we must make sure that we are working in a regular gauge when we Fourier expand.

Let us see how the analysis of regularity works out in the present case. The wordsheet metric is
\begin{empheq}{alignat=7}
	ds^2&=A(\sigma)\left(d\tau^2+d\sigma^2\right)\,,
	&\qquad
	A(\sigma)&=\sinh^2\rho(\sigma)+\sin^2\vartheta_1(\sigma)\,,
\end{empheq}
where the functions $\rho(\sigma)$ and $\vartheta_1(\sigma)$ are defined by
\begin{empheq}{alignat=7}
	\sinh\rho&=\frac{1}{\sinh\sigma}\,,
	&\qquad
	\sin\vartheta_1&=\frac{1}{\cosh\left(\sigma+\sigma_0\right)}\,.
\end{empheq}
The topology is that of a disk with $0<\sigma$ and $\tau\sim\tau+2\pi$. The center of the disk is $\sigma\rightarrow\infty$ where the geometry is flat. To see this, expand near $\sigma=\infty$ to get
\begin{empheq}{alignat=7}
	ds^2&\approx4e^{-2\sigma}\left(1+e^{-2\sigma_0}\right)\left(d\tau^2+d\sigma^2\right)\,.
\end{empheq}
Now let
\begin{empheq}{alignat=7}
	r&=2e^{-\sigma}\sqrt{1+e^{-2\sigma_0}}\,.
\end{empheq}
Then,
\begin{empheq}{alignat=7}
	ds^2&\approx dr^2+r^2d\tau^2\,.
\end{empheq}
This is flat space indeed.

Switching to Cartesian coordinates we have
\begin{empheq}{alignat=7}
	x&=r\cos\tau\,,
	&\qquad
	y&=r\sin\tau\,.
\end{empheq}
The 1-forms transform accordingly:
\begin{empheq}{alignat=7}
	dr&=\frac{xdx+ydy}{\sqrt{x^2+y^2}}\,,
	&\qquad
	d\tau&=\frac{-ydx+xdy}{x^2+y^2}\,,
\end{empheq}
The important fact to remember is that the coordinates $(x,y)$, as well as the 1-forms $dx$ and $dy$ are everywhere well defined. Notice then that neither $dr$ nor $d\tau$ are well defined as $r\rightarrow0$, but the combination $dr^2+r^2d\tau^2$ is. Also, the 1-form $rdr$ is well defined as $r\rightarrow0$ with $rdr\rightarrow0$. In contrast,
\begin{empheq}{alignat=7}
	rd\tau&=\frac{-ydx+xdy}{\sqrt{x^2+y^2}}\,,
\end{empheq}
is ill-defined as $r\rightarrow0$ since the value of the limit depends on the direction in which we approach the origin. This means that only 1-forms involving the combinations
\begin{empheq}{alignat=7}
	r^nd\tau\,,\qquad n\geq2\,,
\end{empheq}
are well defind at $r=0$, where they vanish.

Going back to the worldsheet, the above discussion means that the 1-form $d\tau$ must appear as
\begin{empheq}{alignat=7}
	e^{-n\sigma}d\tau\,,\qquad n\geq2\,,
\end{empheq}
in the gauge fields. In our case we find that
\begin{empheq}{alignat=7}
	\mathcal{A}&=\tanh(2\sigma+\sigma_0)d\tau\approx\left(1-2e^{-4\sigma-2\sigma_0}\right)d\tau\,,
	\\
	\mathcal{B}&=-\frac{1}{2}\tanh(\sigma+\sigma_0)d\tau\approx\left(-\frac{1}{2}+e^{-2\sigma-2\sigma_0}\right)d\tau\,,
\end{empheq}
where we have expanded at large $\sigma$. We see that these gauge fields are not regular at the center of the disk. However, after a gauge transformation we have
\begin{empheq}{alignat=7}
	\mathcal{A}&=\left(\tanh(2\sigma+\sigma_0)-1\right)d\tau\approx-2e^{-4\sigma-2\sigma_0}d\tau\,,
	\\
	\mathcal{B}&=-\frac{1}{2}\left(\tanh(\sigma+\sigma_0)-1\right)d\tau\approx e^{-2\sigma-2\sigma_0}d\tau\,.
\end{empheq}
These gauge fields are  then regular.

\section{Dimensional reduction of spinors}\label{App:Spinors}
Given the symmetries of our problem, the natural way to decompose the ten-dimensional rotations group is
\begin{empheq}{align}
	 SO(10)&\supset\underbrace{SO(2)}_{\gamma}\times\underbrace{SO(2)}_{\rho}\times\underbrace{SO(2)}_{\tau}\times\underbrace{SO(2)}_{\lambda}\times\underbrace{SO(2)}_{\kappa}\,,
\end{empheq}
corresponding to the $(0,1)$, $(2,3)$, $(4,5)$, $(6,7)$ and $(8,9)$ tangent directions, respectively. Under this decomposition, a possible representation of the 10-dimensional gamma matrices is
\begin{empheq}{alignat=5}
	 \Gamma_{\underline{a}}&=\gamma_{\underline{a}}\otimes\mathds{1}\otimes\mathds{1}\otimes\mathds{1}\otimes\mathds{1}\,,
	&\qquad
	\underline{a}&=0,\,1\,,
	\cr
	 \Gamma_{\underline{i}}&=\left(-i\gamma_{\underline{01}}\right)\otimes\rho_{\underline{i}}\otimes\mathds{1}\otimes\mathds{1}\otimes\mathds{1}\,,
	&\qquad
	\underline{i}&=2,\,3\,,
	\cr
	\Gamma_{\underline{i}}&=\left(-i\gamma_{\underline{01}}\right)\otimes\left(-i\rho_{\underline{23}}\right)\otimes\tau_{\underline{i}}\otimes\mathds{1}\otimes\mathds{1}\,,
	&\qquad
	\underline{i}&=4,\,5\,,
	\cr
	\Gamma_{\underline{i}}&=\left(-i\gamma_{\underline{01}}\right)\otimes\left(-i\rho_{\underline{23}}\right)\otimes\left(-i\tau_{\underline{45}}\right)\otimes\lambda_{\underline{i}}\otimes\mathds{1}\,,
	&\qquad
	\underline{i}&=6,\,7\,,
	\cr
	\Gamma_{\underline{i}}&=\left(-i\gamma_{\underline{01}}\right)\otimes\left(-i\rho_{\underline{23}}\right)\otimes\left(-i\tau_{\underline{45}}\right)\otimes\left(-i\lambda_{\underline{67}}\right)\otimes\kappa_{\underline{i}}\,,
	&\qquad
	\underline{i}&=8,\,9\,,
\end{empheq}
where we named the Dirac matrices associated to each factor as displayed above. This basis is tailored for the choice ($\sigma_1$, $\sigma_2$, $\sigma_3$ are Pauli matrices)
\begin{empheq}{alignat=7}
	\gamma_{\underline{0}}&=\rho_{\underline{2}}=\tau_{\underline{4}}=\lambda_{\underline{6}}=\kappa_{\underline{8}}=\sigma_1\,,
	&\qquad
	\gamma_{\underline{1}}&=\rho_{\underline{3}}=\tau_{\underline{5}}=\lambda_{\underline{7}}=\kappa_{\underline{9}}=\sigma_2\,.
\end{empheq}
The chirality operator is then
\begin{empheq}{alignat=7}
	\nonumber
	\Gamma_{11}&\equiv-i\Gamma_{\underline{0123456789}}
	\\
	&=\sigma_3\otimes\sigma_3\otimes\sigma_3\otimes\sigma_3\otimes\sigma_3\,,
\end{empheq}
and the charge conjugation intertwiners $C_{\pm}$ become
\begin{empheq}{alignat=7}
\begin{split}
	C_+&=\Gamma_{\underline{02468}}
	\\
	&=\sigma_1\otimes\left(-i\sigma_2\right)\otimes\sigma_1\otimes\left(-i\sigma_2\right)\otimes\sigma_1
\end{split}\,,
\qquad
\begin{split}
	C_-&=\Gamma_{\underline{13579}}
	\\
	&=\sigma_2\otimes\left(i\sigma_1\right)\otimes\sigma_2\otimes\left(i\sigma_1\right)\otimes\sigma_2
\end{split}\,.
\end{empheq}

A 10-dimensional spinor can be decomposed in terms of 2-dimensional ones as
\begin{empheq}{alignat=7}
	\psi&=\sum_{s_i=\pm}\psi_{s_2s_4s_6s_8}\otimes\eta_{s_2}\otimes\eta_{s_4}\otimes\eta_{s_6}\otimes\eta_{s_8}\,,
\end{empheq}
where
\begin{empheq}{alignat=7}
	\eta_+&=\left(
	\begin{array}{c}
		1
		\\
		0
	\end{array}
	\right)\,,
	&\qquad
	\eta_-&=\left(
	\begin{array}{c}
		0
		\\
		1
	\end{array}
	\right)\,.
\end{empheq}
This provides and explicit projection onto  $\Gamma^{\underline{23}}$, $\Gamma^{\underline{45}}$, $\Gamma^{\underline{67}}$ and $\Gamma^{\underline{89}}$ eigenspaces, with corresponding eigenvalues $-i\alpha\beta\gamma$, $i\alpha$, $i\beta$ and $i\gamma$ which we use in the main body of the text\footnote{The $\kappa$-symmetry fixing in Euclidean language is $i\Gamma_{\underline{01}}\Gamma_{11}\theta=\theta$, where $\Gamma_{11}=-i\Gamma_{\underline{0123456789}}$. This translates to $\Gamma^{\underline{23}}\theta=-i\alpha\beta\gamma\theta$.}.

The Majorana conjugate is
\begin{empheq}{alignat=7}
	\overline{\psi}^M&=\psi^TC_+
	\\
	&= \sum_{s_i=\pm}s_2s_6\:\overline{\psi}^M_{s_2s_4s_6s_8}\otimes\eta_{-s_2}^T\otimes\eta_{-s_4}^T\otimes\eta_{-s_6}^T\otimes\eta_{-s_8}^T\,,
\end{empheq}
with
\begin{empheq}{alignat=7}
	\overline{\psi}^M_{s_2s_4s_6s_8}&\equiv\psi_{s_2s_4s_6s_8}^T\sigma_1\,.
\end{empheq}
Thus, Majorana spinors satisfy
\begin{empheq}{alignat=7}
	\psi^{\dagger}&=\overline{\psi}^M
	&\qquad\Longleftrightarrow\qquad
	s_2s_6\: \overline{\psi}^M_{s_2s_4s_6s_8}&=\psi_{-s_2-s_4-s_6-s_8}^{\dagger}\,.
\end{empheq}

\bibliographystyle{JHEP}
\bibliography{WL-ABJM}

\providecommand{\href}[2]{#2}\begingroup\raggedright\begin{thebibliography}{10}

\bibitem{Pestun:2007rz}
V.~Pestun, {\it {Localization of gauge theory on a four-sphere and
  supersymmetric Wilson loops}},  {\em Commun. Math. Phys.} {\bf 313} (2012)
  71--129, [\href{http://xxx.lanl.gov/abs/0712.2824}{{\tt arXiv:0712.2824}}].

\bibitem{Kapustin:2009kz}
A.~Kapustin, B.~Willett, and I.~Yaakov, {\it {Exact Results for Wilson Loops in
  Superconformal Chern-Simons Theories with Matter}},  {\em JHEP} {\bf 03}
  (2010) 089, [\href{http://xxx.lanl.gov/abs/0909.4559}{{\tt
  arXiv:0909.4559}}].

\bibitem{Drukker:2000ep}
N.~Drukker, D.~J. Gross, and A.~A. Tseytlin, {\it {Green-Schwarz string in
  AdS(5) x S**5: Semiclassical partition function}},  {\em JHEP} {\bf 04}
  (2000) 021, [\href{http://xxx.lanl.gov/abs/hep-th/0001204}{{\tt
  hep-th/0001204}}].

\bibitem{Sakaguchi:2007ea}
M.~Sakaguchi and K.~Yoshida, {\it {A Semiclassical string description of Wilson
  loop with local operators}},  {\em Nucl. Phys.} {\bf B798} (2008) 72--88,
  [\href{http://xxx.lanl.gov/abs/0709.4187}{{\tt arXiv:0709.4187}}].

\bibitem{Kruczenski:2008zk}
M.~Kruczenski and A.~Tirziu, {\it {Matching the circular Wilson loop with dual
  open string solution at 1-loop in strong coupling}},  {\em JHEP} {\bf 05}
  (2008) 064, [\href{http://xxx.lanl.gov/abs/0803.0315}{{\tt
  arXiv:0803.0315}}].

\bibitem{Kristjansen:2012nz}
C.~Kristjansen and Y.~Makeenko, {\it {More about One-Loop Effective Action of
  Open Superstring in $AdS_5\times S^5$}},  {\em JHEP} {\bf 09} (2012) 053,
  [\href{http://xxx.lanl.gov/abs/1206.5660}{{\tt arXiv:1206.5660}}].

\bibitem{Buchbinder:2014nia}
E.~I. Buchbinder and A.~A. Tseytlin, {\it {1/N correction in the D3-brane
  description of a circular Wilson loop at strong coupling}},  {\em Phys. Rev.}
  {\bf D89} (2014), no.~12 126008,
  [\href{http://xxx.lanl.gov/abs/1404.4952}{{\tt arXiv:1404.4952}}].

\bibitem{Forini:2015bgo}
V.~Forini, V.~Giangreco M.~Puletti, L.~Griguolo, D.~Seminara, and E.~Vescovi,
  {\it {Precision calculation of 1/4-BPS Wilson loops in AdS$_5\times S^5$}},
  {\em JHEP} {\bf 02} (2016) 105,
  [\href{http://xxx.lanl.gov/abs/1512.00841}{{\tt arXiv:1512.00841}}].

\bibitem{Faraggi:2016ekd}
A.~Faraggi, L.~A. Pando~Zayas, G.~A. Silva, and D.~Trancanelli, {\it {Toward
  precision holography with supersymmetric Wilson loops}},  {\em JHEP} {\bf 04}
  (2016) 053, [\href{http://xxx.lanl.gov/abs/1601.04708}{{\tt
  arXiv:1601.04708}}].

\bibitem{Forini:2017whz}
V.~Forini, A.~A. Tseytlin, and E.~Vescovi, {\it {Perturbative computation of
  string one-loop corrections to Wilson loop minimal surfaces in AdS$_5 \times$
  S$^5$}},  {\em JHEP} {\bf 03} (2017) 003,
  [\href{http://xxx.lanl.gov/abs/1702.02164}{{\tt arXiv:1702.02164}}].

\bibitem{Cagnazzo:2017sny}
A.~Cagnazzo, D.~Medina-Rincon, and K.~Zarembo, {\it {String corrections to
  circular Wilson loop and anomalies}},
  \href{http://xxx.lanl.gov/abs/1712.07730}{{\tt arXiv:1712.07730}}.

\bibitem{Klemm:2012ii}
A.~Klemm, M.~Marino, M.~Schiereck, and M.~Soroush, {\it
  {Aharony-Bergman-Jafferis-Maldacena Wilson loops in the Fermi gas approach}},
   {\em Z. Naturforsch.} {\bf A68} (2013) 178--209,
  [\href{http://xxx.lanl.gov/abs/1207.0611}{{\tt arXiv:1207.0611}}].

\bibitem{Bianchi:2018bke}
M.~S. Bianchi, L.~Griguolo, A.~Mauri, S.~Penati, and D.~Seminara, {\it {A
  matrix model for the latitude Wilson loop in ABJM theory}},
  \href{http://xxx.lanl.gov/abs/1802.07742}{{\tt arXiv:1802.07742}}.

\bibitem{Aharony:2008ug}
O.~Aharony, O.~Bergman, D.~L. Jafferis, and J.~Maldacena, {\it {N=6
  superconformal Chern-Simons-matter theories, M2-branes and their gravity
  duals}},  {\em JHEP} {\bf 10} (2008) 091,
  [\href{http://xxx.lanl.gov/abs/0806.1218}{{\tt arXiv:0806.1218}}].

\bibitem{Drukker:2008zx}
N.~Drukker, J.~Plefka, and D.~Young, {\it {Wilson loops in 3-dimensional N=6
  supersymmetric Chern-Simons Theory and their string theory duals}},  {\em
  JHEP} {\bf 0811} (2008) 019, [\href{http://xxx.lanl.gov/abs/0809.2787}{{\tt
  arXiv:0809.2787}}].

\bibitem{Rey:2008bh}
S.-J. Rey, T.~Suyama, and S.~Yamaguchi, {\it {Wilson Loops in Superconformal
  Chern-Simons Theory and Fundamental Strings in Anti-de Sitter Supergravity
  Dual}},  {\em JHEP} {\bf 03} (2009) 127,
  [\href{http://xxx.lanl.gov/abs/0809.3786}{{\tt arXiv:0809.3786}}].

\bibitem{Chen:2008bp}
B.~Chen and J.-B. Wu, {\it {Supersymmetric Wilson Loops in N=6 Super
  Chern-Simons-matter theory}},  {\em Nucl. Phys.} {\bf B825} (2010) 38--51,
  [\href{http://xxx.lanl.gov/abs/0809.2863}{{\tt arXiv:0809.2863}}].

\bibitem{Marino:2009jd}
M.~Marino and P.~Putrov, {\it {Exact Results in ABJM Theory from Topological
  Strings}},  {\em JHEP} {\bf 06} (2010) 011,
  [\href{http://xxx.lanl.gov/abs/0912.3074}{{\tt arXiv:0912.3074}}].

\bibitem{Drukker:2010nc}
N.~Drukker, M.~Marino, and P.~Putrov, {\it {From weak to strong coupling in
  ABJM theory}},  {\em Commun. Math. Phys.} {\bf 306} (2011) 511--563,
  [\href{http://xxx.lanl.gov/abs/1007.3837}{{\tt arXiv:1007.3837}}].

\bibitem{Drukker:2009hy}
N.~Drukker and D.~Trancanelli, {\it {A Supermatrix model for N=6 super
  Chern-Simons-matter theory}},  {\em JHEP} {\bf 02} (2010) 058,
  [\href{http://xxx.lanl.gov/abs/0912.3006}{{\tt arXiv:0912.3006}}].

\bibitem{Fuji:2011km}
H.~Fuji, S.~Hirano, and S.~Moriyama, {\it {Summing Up All Genus Free Energy of
  ABJM Matrix Model}},  {\em JHEP} {\bf 08} (2011) 001,
  [\href{http://xxx.lanl.gov/abs/1106.4631}{{\tt arXiv:1106.4631}}].

\bibitem{Marino:2011eh}
M.~Marino and P.~Putrov, {\it {ABJM theory as a Fermi gas}},  {\em J. Stat.
  Mech.} {\bf 1203} (2012) P03001,
  [\href{http://xxx.lanl.gov/abs/1110.4066}{{\tt arXiv:1110.4066}}].

\bibitem{Cardinali:2012ru}
V.~Cardinali, L.~Griguolo, G.~Martelloni, and D.~Seminara, {\it {New
  supersymmetric Wilson loops in ABJ(M) theories}},  {\em Phys. Lett.} {\bf
  B718} (2012) 615--619, [\href{http://xxx.lanl.gov/abs/1209.4032}{{\tt
  arXiv:1209.4032}}].

\bibitem{Bianchi:2014laa}
M.~S. Bianchi, L.~Griguolo, M.~Leoni, S.~Penati, and D.~Seminara, {\it {BPS
  Wilson loops and Bremsstrahlung function in ABJ(M): a two loop analysis}},
  {\em JHEP} {\bf 06} (2014) 123,
  [\href{http://xxx.lanl.gov/abs/1402.4128}{{\tt arXiv:1402.4128}}].

\bibitem{Bonini:2016fnc}
M.~Bonini, L.~Griguolo, M.~Preti, and D.~Seminara, {\it {Surprises from the
  resummation of ladders in the ABJ(M) cusp anomalous dimension}},  {\em JHEP}
  {\bf 05} (2016) 180, [\href{http://xxx.lanl.gov/abs/1603.00541}{{\tt
  arXiv:1603.00541}}].

\bibitem{Bianchi:2017svd}
M.~S. Bianchi, L.~Griguolo, A.~Mauri, S.~Penati, M.~Preti, and D.~Seminara,
  {\it {Towards the exact Bremsstrahlung function of ABJM theory}},  {\em JHEP}
  {\bf 08} (2017) 022, [\href{http://xxx.lanl.gov/abs/1705.10780}{{\tt
  arXiv:1705.10780}}].

\bibitem{Bianchi:2017ozk}
L.~Bianchi, L.~Griguolo, M.~Preti, and D.~Seminara, {\it {Wilson lines as
  superconformal defects in ABJM theory: a formula for the emitted radiation}},
   {\em JHEP} {\bf 10} (2017) 050,
  [\href{http://xxx.lanl.gov/abs/1706.06590}{{\tt arXiv:1706.06590}}].

\bibitem{Bianchi:2018scb}
L.~Bianchi, M.~Preti, and E.~Vescovi, {\it {Exact Bremsstrahlung functions in
  ABJM theory}},  \href{http://xxx.lanl.gov/abs/1802.07726}{{\tt
  arXiv:1802.07726}}.

\bibitem{Correa:2014aga}
D.~H. Correa, J.~Aguilera-Damia, and G.~A. Silva, {\it {Strings in $AdS_4
  \times \mathbb{CP}^{3}$ Wilson loops in $\mathcal N=$6 super
  Chern-Simons-matter and bremsstrahlung functions}},  {\em JHEP} {\bf 06}
  (2014) 139, [\href{http://xxx.lanl.gov/abs/1405.1396}{{\tt
  arXiv:1405.1396}}].

\bibitem{Sakaguchi:2010dg}
M.~Sakaguchi, H.~Shin, and K.~Yoshida, {\it {Semiclassical Analysis of M2-brane
  in $AdS_4 x S^7 / Z_k$}},  {\em JHEP} {\bf 1012} (2010) 012,
  [\href{http://xxx.lanl.gov/abs/1007.3354}{{\tt arXiv:1007.3354}}].

\bibitem{Kim:2012nd}
H.~Kim, N.~Kim, and J.~H. Lee, {\it {One-loop corrections to holographic Wilson
  loop in AdS4xCP3}},  {\em J.Korean Phys.Soc.} {\bf 61} (2012) 713--719,
  [\href{http://xxx.lanl.gov/abs/1203.6343}{{\tt arXiv:1203.6343}}].

\bibitem{Camporesi:1994ga}
R.~Camporesi and A.~Higuchi, {\it {Spectral functions and zeta functions in
  hyperbolic spaces}},  {\em J. Math. Phys.} {\bf 35} (1994) 4217--4246.

\bibitem{Camporesi:1995fb}
R.~Camporesi and A.~Higuchi, {\it {On the Eigen functions of the Dirac operator
  on spheres and real hyperbolic spaces}},  {\em J. Geom. Phys.} {\bf 20}
  (1996) 1--18, [\href{http://xxx.lanl.gov/abs/gr-qc/9505009}{{\tt
  gr-qc/9505009}}].

\bibitem{Aguilera-Damia:2018rjb}
J.~Aguilera-Damia, A.~Faraggi, L.~Pando~Zayas, V.~Rathee, and G.~A. Silva, {\it
  {Functional Determinants of Radial Operators in $AdS_2$}},
  \href{http://xxx.lanl.gov/abs/1802.06789}{{\tt arXiv:1802.06789}}.

\bibitem{Aguilera-Damia:2018twq}
J.~Aguilera-Damia, A.~Faraggi, L.~A. Pando~Zayas, V.~Rathee, and G.~A. Silva,
  {\it {Zeta-function Regularization of Holographic Wilson Loops}},
  \href{http://xxx.lanl.gov/abs/1802.03016}{{\tt arXiv:1802.03016}}.

\bibitem{Medina-Rincon:2018wjs}
D.~Medina-Rincon, A.~A. Tseytlin, and K.~Zarembo, {\it {Precision matching of
  circular Wilson loops and strings in AdS(5)xS(5)}},
  \href{http://xxx.lanl.gov/abs/1804.08925}{{\tt arXiv:1804.08925}}.

\bibitem{Muck:2016hda}
W.~M\"uck, L.~A. Pando~Zayas, and V.~Rathee, {\it {Spectra of Certain
  Holographic ABJM Wilson Loops in Higher Rank Representations}},  {\em JHEP}
  {\bf 11} (2016) 113, [\href{http://xxx.lanl.gov/abs/1609.06930}{{\tt
  arXiv:1609.06930}}].

\bibitem{Cookmeyer:2016dln}
J.~Cookmeyer, J.~T. Liu, and L.~A. Pando~Zayas, {\it {Higher Rank ABJM Wilson
  Loops from Matrix Models}},  {\em JHEP} {\bf 11} (2016) 121,
  [\href{http://xxx.lanl.gov/abs/1609.08165}{{\tt arXiv:1609.08165}}].

\end{thebibliography}\endgroup

\end{document}